\documentclass[a4paper,12pt]{article}

\usepackage{indentfirst}
\usepackage{amsmath, amssymb, amscd, amsthm, amsfonts}
\usepackage{mathabx,mathtools}
\usepackage{bm}
\usepackage{graphicx}
\graphicspath{{fig}}
\usepackage{enumitem}
\usepackage{natbib}
\usepackage{url}
\usepackage{hyperref}
\usepackage{multirow}
\usepackage{caption}
\usepackage{color}
\usepackage{xcolor}
\usepackage[margin=1in]{geometry}
\usepackage{hhline}
\usepackage{appendix}
\usepackage{soul}
\usepackage{xr}
\usepackage{algorithm,algorithmic}

\newtheorem{theorem}{Theorem}

\newtheorem{proposition}{Proposition}
\newtheorem{remark}{Remark}

\newtheorem{assumption}{Assumption}

\allowdisplaybreaks[4]

\makeatletter
\newenvironment{breakablealgorithm}
  {% \begin{breakablealgorithm}
    \begin{center}
      \refstepcounter{algorithm}% New algorithm
      \hrule height.8pt depth0pt \kern2pt% \@fs@pre for \@fs@ruled
      \parskip 0pt
      \renewcommand{\caption}[2][\relax]{% Make a new \caption
        {\raggedright\textbf{\fname@algorithm~\thealgorithm} ##2\par}%
        \ifx\relax##1\relax % #1 is \relax
          \addcontentsline{loa}{algorithm}{\protect\numberline{\thealgorithm}##2}%
        \else % #1 is not \relax
          \addcontentsline{loa}{algorithm}{\protect\numberline{\thealgorithm}##1}%
        \fi
        \kern2pt\hrule\kern2pt
     }
  }
  {% \end{breakablealgorithm}
     \kern2pt\hrule\relax% \@fs@post for \@fs@ruled
   \end{center}
  }
\makeatother

\newcommand{\trans}{^{\scriptscriptstyle \sf T}}
\newcommand{\pr}{\mathsf{P}}

\newcommand{\calI}{\mathcal{I}}

\newcommand{\bbR}{\mathbb{R}}

\begin{document}

\title{Optimal Integrative Estimation for Distributed Precision Matrices with Heterogeneity Adjustment}

\author{
Yinrui Sun\thanks{Department of Statistics and Data Science, Fudan University.} 
\and Yin Xia\thanks{Department of Statistics and Data Science, Fudan University.}
}

\date{ }

\maketitle

\baselineskip=20pt

\begin{abstract}\normalsize
Distributed learning offers a practical solution for the integrative analysis of multi-source datasets, especially under privacy or communication constraints.
However, addressing prospective distributional heterogeneity and ensuring communication efficiency pose significant challenges on distributed statistical analysis.
In this article, we focus on integrative estimation of distributed heterogeneous precision matrices, a crucial task related to joint precision matrix estimation where computation-efficient algorithms and statistical optimality theories are still underdeveloped.
To tackle these challenges, we introduce a novel HEterogeneity-adjusted Aggregating and Thresholding (HEAT) approach for distributed integrative estimation.
HEAT is designed to be both communication- and computation-efficient, and we demonstrate its statistical optimality by establishing the convergence rates and the corresponding minimax lower bounds under various integrative losses.
To enhance the optimality of HEAT, we further propose an iterative HEAT (IteHEAT) approach. 
By iteratively refining the higher-order errors of HEAT estimators through multi-round communications, IteHEAT achieves geometric contraction rates of convergence.
Extensive simulations and real data applications validate the numerical performance of HEAT and IteHEAT methods.
\end{abstract}


\newpage

%%%%%%%%%%%%%%%%%%%%%%%%%%%%%%%%%%%%%%%%%%%%%%%%%%%%%%
%%%%%%%%%%%%%%%%%%%%%%%%%%%%%%%%%%%%%%%%%%%%%%%%%%%%%%
\section{Introduction}\label{sec:intro}
%%%%%%%%%%%%%%%%%%%%%%%%%%%%%%%%%%%%%%%%%%%%%%%%%%%%%%
%%%%%%%%%%%%%%%%%%%%%%%%%%%%%%%%%%%%%%%%%%%%%%%%%%%%%%

The proliferation of high-dimensional data presents numerous opportunities for statistical learning but can lead to less accurate analyses when applying targeted algorithms to a single local dataset with limited sample size. 
Fortunately, advancements in big data collection, storage, and transmission capabilities have made it feasible to access multiple datasets containing similar information.
These datasets may originate from diverse populations, institutions and geographical locations, such as neuroimaging data \citep{bellec2017neuro}, electronic health records \citep{kohane2021what}, and genetic data \citep{zhou2022global}.
By leveraging the inherent similarities among these distributed data, integrative analysis holds significant promise for enhancing statistical accuracy. 

Nevertheless, synthesizing information from multiple sources is highly challenging.
First, the prospective between-study heterogeneity complicates data integration.
Such heterogeneity may manifest in various forms, including variations in genotype-phenotype associations \citep{frazierwood2013genetic} and the differences in genetic variants across diverse ethnic populations \citep{gurdasani2019genomics}.
Additionally, variability arising from inconsistent data collection and measurement across institutions further contributes to data heterogeneity \citep{kohane2021what}.
Second, due to privacy \citep{arellano2018privacy,kohane2021what} or communication concerns, individual raw data often cannot be shared across different locations.
Instead, only summary or intermediate statistics can be communicated among different sources \citep{duan2020learning,zhou2022global}.
This constraint on data sharing may negatively impact the learning accuracy, exacerbating the challenges of data integration.

In this article, we focus on the distributed integrative estimation of high-dimensional precision matrices by taking into account the aforementioned practical challenges.
The precision matrix, defined as the inverse of the covariance matrix, is central to various statistical models and plays a pivotal role in mathematical statistics.
Moreover, the sparsity pattern of a precision matrix serves as a valuable tool for characterizing conditional associations between covariates, and is therefore widely employed in many scientific applications such as brain connectivity \citep{fransson2008precuneus} and genetic network analysis \citep{de2004discovery}.
Additionally, integrative analysis of the multi-source datasets significantly enhances the association study, while the patterns of heterogeneity facilitate differential analysis \citep{min2012coexpression,bellec2017neuro,shojaie2021differential}.
Consequently, there is a critical need to focus on the distributed setup and develop communication-efficient algorithms for estimating heterogeneous precision matrices with solid statistical guarantees.

%%%%%%%%%%%%%%%%%%%%%%%%%%%%%%%%%%%%%%%%%%%%%%%%%%%%%%
\subsection{Related Works}\label{sec:literature}
%%%%%%%%%%%%%%%%%%%%%%%%%%%%%%%%%%%%%%%%%%%%%%%%%%%%%%

In recent decades, significant progress has been made in the field of high-dimensional precision matrix estimation under sparsity, including penalized maximum likelihood procedures \citep{yuan2007model,rothman2008sparse,lam2009sparsistency,ravikumar2011high}, penalized linear regression approaches \citep{meinshausen2006high,sun2013sparse,ren2015asymptotic,liu2015fast}, and regularized estimating equation methods \citep{yuan2010high,cai2011clime,cai2016estimating}.
The non-negligible bias induced by the regularization in these methods has prompted numerous debiasing strategies aimed at statistical inference on the entries of precision matrices \citep{liu2013gaussian,jankova2015confidence,jankova2017honest,
ning2017general,neykov2018unified,chang2018confidence}.

To enhance the estimation accuracy of the precision matrix within a single dataset, a rich amount of joint estimation proposals have emerged.
Notable examples include $\ell_1/\ell_q$ penalized procedures, e.g., $q=2$  \citep{danaher2014joint,ren2019tuning} or $q=\infty$ \citep{honorio2010multitask,cai2016joint}, non-convex penalized procedures \citep{guo2011joint,chun2015gene}, and decomposition-based methods that assume a structure involving a common matrix and individual matrices \citep{hara2013learning,lee2015joint}; see a thorough survey in \citet{tsai2022joint} and many other references therein.
Nevertheless, the aforementioned joint estimation methods typically reply on solving highly non-smooth regularized optimization problems, leading to computationally intensive procedures.
Additionally, due to the intrinsic complexity of precision matrix estimation, statistical optimality theories for these methods remain unknown.
%{\magenta Moreover, it is also worth noting that although almost all of these methods depend only on the sample covariances, which can serve as the summary statistics in the distributed settings, the simultaneous storage of multiple large covariances significantly burdens the memory of working machine.
%}{\blue [I am still confused about this claim. According to the current writing of the paper, HEAT also requires the storage of multiple precision matrix estimators (we did not mention we can just send one entry at a time; though the supplement mentioned briefly that we can compute $L_m$ at local site, it is not detailed in the paper), so the memory burden is the same as the joint estimation?]}

In a similar context to joint estimation, there is a substantial body of literature focusing on distributed analysis under distribution homogeneity. 
Examples include the averaging methods \citep{zhang2015divide,lee2017communication,battey2018distributed,fan2019distributed,
dobriban2020wonder,dobriban2021distributed,lv2022debiased,tu2024distributedave}, and the surrogate loss function methods \citep{wang2017efficient,jordan2019communication} with applications in distributed optimization \citep{fan2023communication}, simultaneous inference \citep{yu2022distributed}, non-smooth analysis \citep{tan2022communication} and semi-supervised estimation \citep{tu2024distributed}.
These methods achieve communication-efficiency in the sense that the dimensions of the transmitted statistics are proportional to those of target parameters.
Besides, averaging aggregations under communication constraints are developed for estimation \citep{garg2014communication,braverman2016communication,zhu2018distributed,
szabo2020adaptive,cai2022distributed,acharya2023unified,acharya2024optimal} and testing \citep{szabo2022optimal,szabo2023optimal}.
In addition, \citet{chen2019quantile} smooths the non-smooth objective and solves it by transmitting statistics with dimensions that are quadratic in the parameter dimension; see also \citet{wang2019distributed,chen2024distributed} for further applications.
On the other hand, when model distributions are heterogeneous while target parameters remain homogeneous across sites, distributed analysis can be achieved by averaging \citep{zhao2016partially,gu2023distributed,chen2025distributed} or surrogate estimating equations \citep{duan2022heterogeneity}.

Despite these advancements in homogeneous settings, statistical analysis for handling distributed heterogeneous parameters remain relatively rare. 
For instance, \citet{maity2022meta} identifies and estimates the representative global parameter across sites, while \citet{guo2025robust} focuses on robust estimation of the majority parameter. 
Both approaches requires only local parameter estimators from the distributed sites.
\citet{liu2025robust} addresses heterogeneity by fusion regularization and solves the regression problem through a first-order distributed optimization algorithm. 
\citet{cai2022individual} introduces a quadratic surrogate loss function and transmits the first-order gradients and second-order Hessians to jointly estimate the distributed regression coefficients, with further extensions to multiple testing \citep{liu2021integrative} and transfer learning \citep{li2023targeting}.
Nonetheless, developing communication-efficient and statistically optimal methods for integrative analysis in the presence of heterogeneity remains an active area of research.

%%%%%%%%%%%%%%%%%%%%%%%%%%%%%%%%%%%%%%%%%%%%%%%%%%%%%%
\subsection{Contributions}
%%%%%%%%%%%%%%%%%%%%%%%%%%%%%%%%%%%%%%%%%%%%%%%%%%%%%%

In this article, we aim to develop distributed integrative estimation algorithms for heterogeneous precision matrices, ensuring both communication and computation efficiency along with statistically provable optimality.
A novel method, HEterogeneity-adjusted Aggregating and Thresholding (HEAT), is proposed to achieve these goals effectively.
Methodologically, the distributed HEAT algorithm consists of two main steps. 
First, an individual local estimation is derived at each local site, and summary statistics are then transmitted to the central site.
Second, without any individual-level information, an integration step is conducted at the central site by performing HEAT.
Theoretically, we drive the estimation rates of convergence for HEAT over a class of heterogeneous precision matrices under various norms of the introduced integrative loss matrices.
Moreover, we demonstrate the statistical optimality of HEAT by establishing the matched minimax lower bounds for integrative estimation. 
Finally, to enhance the optimality, we further propose an iterative HEAT (IteHEAT) method that iteratively refines the higher-order errors of HEAT through multi-round communications between local sites and the central site.
A contraction theory is derived for IteHEAT, demonstrating the geometric vanishing rates of higher-order errors and thereby indicating improved estimation accuracy.

The newly established methodologies and theories in this article make several valuable contributions to both distributed statistical learning and integrative precision matrix analysis.
First, the proposed HEAT and IteHEAT methods introduce novel methodological advances in distributed algorithms under heterogeneity. 
In comparison to existing communication-efficient algorithms developed under homogeneity, our approaches demonstrate flexibility and applicability across distributional heterogeneity scenarios. 
Additionally, our methods provide communication advantages over distributed algorithms designed for heterogeneous settings, such as \citet{cai2022individual}.
Second, the proposed IteHEAT algorithm provides insights into the statistical refinement of higher-order errors and demonstrates potential for improving estimation accuracy in integrative analysis.
Third, the proposed methods complement the field of multi-task learning for precision matrices.
Unlike the typical joint estimation approaches that rely on solving highly non-smooth regularized optimizations, HEAT and IteHEAT are optimization-free for integration, ensuring computational efficiency.
Fourth, we establish the statistical optimality theories of HEAT and IteHEAT by deriving the rates of convergence and the corresponding minimax lower bounds.
To the best of our knowledge, the current minimax theories are the first among the extensive literature on joint precision matrix estimation. 
The introduced lower bound techniques and constructions of least favorable matrices are of independent interest and make valuable contributions to the field.

\subsection{Organization}
The rest of the article is organized as follows.
Section \ref{sec:notation_formulation} introduces the problem framework.
Section \ref{sec:method} proposes the HEAT method, including the individual estimation step at the local sites and the integration step at the central site.
The theoretical results of HEAT are established in Section \ref{sec:theory_up}, and the minimax lower bounds are derived in Section \ref{sec:theory_low}.
Section \ref{sec:iteration} introduces the refined IteHEAT method and provides the corresponding theoretical analysis.
The numerical performance of the proposed methods is evaluated through simulations in Section \ref{sec:simulation}, followed by a real data analysis in Section \ref{sec:real_data}.
Section \ref{sec:discuss} concludes the article with discussions. 
All technical proofs and additional results are collected in the Supplementary Material.

%%%%%%%%%%%%%%%%%%%%%%%%%%%%%%%%%%%%%%%%%%%%%%%%%%%%%%
%%%%%%%%%%%%%%%%%%%%%%%%%%%%%%%%%%%%%%%%%%%%%%%%%%%%%%
\section{Notations and Problem Setup}\label{sec:notation_formulation}
%%%%%%%%%%%%%%%%%%%%%%%%%%%%%%%%%%%%%%%%%%%%%%%%%%%%%%
%%%%%%%%%%%%%%%%%%%%%%%%%%%%%%%%%%%%%%%%%%%%%%%%%%%%%%

\subsection{Notations}\label{sec:notation}

For an integer $k>0$, let $[k] = \{1,2,\ldots,k\}$; denote by $0_k$ and $1_k$ the vectors of zeros and ones with length $k$, respectively.
For a set $\mathcal{A}$, define $|\mathcal{A}|$ to be its cardinality.
For a vector $a\in\bbR^n$ and a subset $\mathcal{A} \subset [n]$, denote by $a_\mathcal{A} = \left(a_i\right)_{i\in \mathcal{A}} \in \bbR^{|\mathcal{A}|}$ a subvector of $a$.
Let $\mathcal{A}^c = [n] \setminus \mathcal{A}$, and define $a_{-\mathcal{A}} = a_{\mathcal{A}^c} \in \bbR^{n-|\mathcal{A}|}$, and $a_{-j} = a_{-\{j\}}$.
For a matrix $A = \left(A_{i,j}\right)_{i\in[n_1],j\in[n_2]} \in \bbR^{n_1 \times n_2}$ and subsets $\mathcal{A}_1 \subset [n_1],~\mathcal{A}_2 \subset [n_2]$, let $A_{\mathcal{A}_1,\mathcal{A}_2} = \left(A_{i,j}\right)_{i\in\mathcal{A}_1,j\in\mathcal{A}_2} \in \bbR^{|\mathcal{A}_1|\times |\mathcal{A}_2|}$ be a submatrix of $A$.
For $j\in[n_2]$, define $A_{\cdot,j} = A_{[n_1],\{j\}}$ and $A_{\cdot,-j} = A_{[n_1],\{j\}^c}$.
Define $|A|_\infty = \max_{i\in[n_1], j\in[n_2]} |A_{i,j}|$ to be the entry-wise maximum norm, and $\| A \|_l = \sup_{x\in \bbR^{n_2}\setminus \{0_{n_2}\}  } \| Ax \|_l / \|x\|_l$ to be the matrix $l$-norm with $l \in [1,\infty]$.
Denote by $S_{+}^n$ the set of $n \times n$ symmetric positive definite matrices.

For two positive deterministic sequences $\{a_n\}$ and $\{b_n\}$, write $a_n \lesssim b_n$ if there exists a constant $C>0$ such that $a_n\leq Cb_n$ for all sufficiently large $n$, and $a_{n} \asymp b_{n}$ if $a_n \lesssim b_n$ and $b_n \lesssim a_n$.
For two sequences of positive random variables $\{X_n\}$ and $\{Y_n\}$, write $X_n \lesssim_\pr Y_n$ if $X_n \lesssim Y_n$ holds with probability tending to 1 as $n\rightarrow\infty$.
For a random variable $X$, define $\|X\|_{e_k} = \inf\{ t\geq 0: \mathbb{E} e_k(|X| / t) \leq 1 \}$, where $e_k(x) = \exp(x^k) - 1$ for $k\geq 1$; 
$\|X\|_{e_2}$ ($\|X\|_{e_1}$) is called the sub-Gaussian (sub-exponential) norm of $X$,
and $X$ is sub-Gaussian (sub-exponential) if $\|X\|_{e_2}$ ($\|X\|_{e_1}$) is finite.
A random vector $X \in \bbR^n$ is sub-Gaussian if $a\trans X$ is sub-Gaussian for any deterministic vector $a\in \bbR^n$, and its sub-Gaussian norm is defined as $ \| X \|_{e_2} := \sup_{a\in \bbR^n: \|a\|_2 = 1} \| a\trans X \|_{e_2}$.

Let $M$ be the total number of distributed datasets. 
Define $a^{(\cdot)} = \left( a^{(1)}, \ldots, a^{(M)} \right)\trans \in \bbR^{M}$ for $a^{(1)}, \ldots, a^{(M)} \in \bbR$, and $a^{(\cdot,t)} = \left( a^{(1,t)}, \ldots, a^{(M,t)} \right)\trans \in \bbR^{M}$ for $a^{(1,t)}, \ldots, a^{(M,t)} \in \bbR$.
For matrices $A^{(1)}, \ldots, A^{(M)} \in \bbR^{n_1 \times n_2}$, define $A^{(\cdot)} = \left( A^{(1)}, \ldots, A^{(M)} \right) \in (\bbR^{n_1 \times n_2})^{\otimes M}$; for matrices $A^{(1,t)}, \ldots, A^{(M,t)} \in \bbR^{n_1 \times n_2}$, define $A^{(\cdot,t)} = \left( A^{(1,t)}, \ldots, A^{(M,t)} \right) \in (\bbR^{n_1 \times n_2})^{\otimes M}$.
Let $C,C_0,C_1,c,c_0,c_1,\ldots$ be universal positive constants that may differ from place to place.

%%%%%%%%%%%%%%%%%%%%%%%%%%%%%%%%%%%%%%%%%%%%%%%%%%%%%%
\subsection{Problem Setup}\label{sec:formulation}
%%%%%%%%%%%%%%%%%%%%%%%%%%%%%%%%%%%%%%%%%%%%%%%%%%%%%%

Recall that, we focus on the distributed scenarios where multiple datasets are stored separately in $M$ different sites.
Denote those local datasets by $\{ X^{(m)} \}_{m \in[M]}$, where $X^{(m)} = (X_1^{(m)}, \ldots, X_{n_m}^{(m)} )\trans \in \bbR^{n_m \times p}$ and $X_i^{(m)}\in\bbR^{p}, i\in [n_m]$ are independently and identically distributed (i.i.d.) with population mean $\mu^{(m)}\in\bbR^{p}$ and covariance $\Sigma^{(m)}\in S_{+}^{p}$. 
Let $N=\sum_{m=1}^M n_m$ be the total number of samples, and denote by $n = \min_{m\in[M]} n_m$ and $n_{\max} = \max_{m\in[M]} n_m$ the minimal and maximal individual sample sizes, respectively.
Throughout, we consider the high-dimensional regime where $(n,p)\rightarrow \infty$, and assume that there exists a constant $c>0$ such that $M \lesssim p^c$, which is mild under high-dimensionality.

Due to potential privacy or communication constraints on data sharing, direct merging of individual raw data is typically prohibited. 
Instead, only summary or intermediate statistics can be communicated from local sites to a central machine, which presents significant challenges for integration and analysis. 
In this context, the parameters of interest in this article are the precision matrices, defined as
$$
\Omega^{(m)} = ( \Omega_{j,k}^{(m)} )_{j,k\in[p]} := ( \Sigma^{(m)} )^{-1}, m\in [M],
$$
and we focus on their estimation under sparsity.
Besides, the datasets from different sources may exhibit heterogeneous distributions, implying potentially different precision matrices across local sites. 
To accommodate such heterogeneity, for each $k\in[p]$, we decompose the indices of the non-zero components in the $k$-th columns into two sets:
$$
S_{1,k} = \{ j \in [p]: \Omega_{j, k}^{(1)} = \cdots = \Omega_{j, k}^{(M)} \neq 0 \} ,
$$ 
$$
S_{2,k} = \{ j \in [p]: \text{ not all } \Omega_{j, k}^{(1)}, \cdots, \Omega_{j, k}^{(M)} \text{ are equal} \} .
$$
The entries in $S_{1,k}$ suggest a common non-zero structure, while those in $S_{2,k}$ indicate the heterogeneity.
These implications yield the following parameter space:
\begin{equation}\label{eq:para_space_mot}
\Theta(s_1,s_2) = \left\{ \Omega^{(\cdot)} \in (S_{+}^p)^{\otimes M}: \max_{k\in[p]} |S_{1,k}| \leq s_1, \max_{k\in[p]}|S_{2,k}| \leq s_2 \right\}.
\end{equation}
Our goal is to develop distributed integrative estimation procedures 
for the precision matrices within the space \eqref{eq:para_space_mot}, while ensuring communication efficiency and statistical guarantees. 

It is worth highlighting that, $s_2 = 0$ corresponds to complete homogeneity $\Omega^{(1)} = \cdots = \Omega^{(M)}$, while $s_1 = 0$ indicates the group sparsity \citep{yuan2006model,lounici2011oracle} that has been widely considered in joint precision matrix estimations \citep{danaher2014joint,cai2016joint,ren2019tuning}.
Therefore, the parameter space \eqref{eq:para_space_mot} is flexible and can effectively model the similarity and heterogeneity.
However, this flexibility, combined with the computational and statistical challenges inherent in precision matrix estimation, further amplifies the difficulty of distributed integrative analysis.

%%%%%%%%%%%%%%%%%%%%%%%%%%%%%%%%%%%%%%%%%%%%%%%%%%%%%%
%%%%%%%%%%%%%%%%%%%%%%%%%%%%%%%%%%%%%%%%%%%%%%%%%%%%%%
\section{Methodology}\label{sec:method}
%%%%%%%%%%%%%%%%%%%%%%%%%%%%%%%%%%%%%%%%%%%%%%%%%%%%%%
%%%%%%%%%%%%%%%%%%%%%%%%%%%%%%%%%%%%%%%%%%%%%%%%%%%%%%

In this section, we propose a distributed integrative algorithm for heterogeneous precision matrices estimation.
The algorithm consists of two main steps.
First, each local site computes its individual estimate and then transmits the local summary statistics to the central site. 
Second, the central site performs the HEterogeneity-adjusted Aggregating and Thresholding (HEAT) step to integrate the outputs from the local sites.
The proposed method is outlined in Algorithm \ref{alg:alg_main}, and the detailed steps are presented in Sections \ref{sec:method_local} and \ref{sec:method_heat}.

\begin{breakablealgorithm}
\caption{Distributed Estimation for Heterogeneous Precision Matrices}
\label{alg:alg_main}

\begin{algorithmic}

~

\textbf{Step 1. Individual Local Estimation:}

For $m \in [M]$, at the $m$-th site, do:

\begin{enumerate}[nosep]

\item Lasso regression: calculate $\widehat{\gamma}_j^{(m) } = \arg\min_{ \gamma\in\bbR^{p-1} } \left\{ \frac{1}{2 n_m} \| \widecheck{X}_{\cdot, j}^{(m)} - \widecheck{X}_{\cdot,-j}^{(m)} \gamma \|_2^2 + \lambda_j^{(m)} \left\| \gamma \right\|_1\right\}$, for $j\in[p]$, where $\widecheck{X}_{i,j}^{(m)} := X_{i,j}^{(m)} - \widebar{X}_{j}^{(m)}$ and $\widebar{X}^{(m)} = \frac{1}{n_m} \sum_{i=1}^{n_m} X_i^{(m)}$.

\item Regularized estimation: obtain $\widehat{\Omega}^{(m)} = ( \widehat{\Omega}_{j,k}^{(m)} )_{j,k\in[p]}$ via \eqref{eq:Omega_jj_est}-\eqref{eq:Omega_j_est} based on $( \widehat{\gamma}_j^{(m)} )_{j\in[p]}$.

\item Debiasing: compute $\widebar{\Omega}^{(m)} = \widehat{\Omega}^{(m)} + (\widehat{\Omega}^{(m)})\trans - ( \widehat{\Omega}^{(m)} )\trans \widehat{\Sigma}^{(m)} \widehat{\Omega}^{(m)}$, where $ \widehat{\Sigma}^{(m)} = n^{-1}_m\sum_{i=1}^{n_m} (X_i^{(m)}-\widebar{X}^{(m)}) (X_i^{(m)}-\widebar{X}^{(m)})\trans$.

\item Transmit the quantities $n_m$ and $\widebar{\Omega}^{(m)}$ to the central site.

\item (Optional) Variance estimation: compute $( \widehat{v}_{j,k}^{(m)} )_{j,k\in[p]}$ via \eqref{eq:variance_est}, and transmit $( \widehat{v}_{j,k}^{(m)} )_{j,k\in[p]}$ to the central site.

\end{enumerate}

\textbf{Step 2. Heterogeneity-Adjusted Aggregating and Thresholding:}

At the central site, do:

\begin{enumerate}[nosep]

% \item Shrinkage threshold: compute the $\lambda_{1,j,k}$ and $ \lambda_{1,j,k}$ based on the $\{n_m\}_{m\in[M]}$ and $( \widehat{v}_{j,k}^{(m)} )_{m\in[M]}$ as well as the equations \eqref{eq:lambda_1jk}-\eqref{eq:lambda_2jk} in Section \ref{sec:theory_up} for $j,k\in[p]$.

\item Integrative estimation via HEAT: 
$$
\widetilde{\Omega}^{(m)} = \widetilde{\Gamma} + \widetilde{\Lambda}^{(m)} = ( \widetilde{\Gamma}_{j,k} )_{j,k\in[p]} + ( \widetilde{\Lambda}_{j,k}^{(m)} )_{j,k\in[p]}, ~ m\in[M] ,
$$
with
$
\widetilde{\Gamma}_{j, k} = T_{1, \lambda_{1, j, k} }( \widebar{\Omega}_{j,k} )$, $\widetilde{\Lambda}_{j, k}^{(\cdot)} = T_{2, \lambda_{2, j, k}} ( \widebar{\Omega}_{j, k}^{(\cdot)} - \widebar{\Omega}_{j,k} 1_M ),
$
and $\widebar{\Omega}_{j, k} = \frac{1}{N}\sum_{m = 1}^M n_m \widebar{\Omega}_{j, k}^{(m)}$, where $T_{1, \lambda}: \bbR \rightarrow \bbR$, $T_{2, \lambda}: \bbR^M \rightarrow \bbR^M$ are thresholding functions, and $\lambda_{1,j,k},\lambda_{2,j,k} \geq 0$ are shrinkage levels.

\item Send $\widetilde{\Omega}^{(m)}$ back to the $m$-th local site for $m\in[M]$.

\end{enumerate}

\end{algorithmic}
\end{breakablealgorithm}

\subsection{Individual Local Estimation}\label{sec:method_local}

In the first step, we aim to obtain an individual estimation for $\Omega^{(m)}$ based on the data $X^{(m)}$ at the $m$-th local site, for each $m\in [M]$.
Define the projection coefficients by
\begin{equation}\label{eq:gamma_j_def}
\gamma_j^{(m)} = \arg\min_{ \gamma \in \bbR^{p-1} } \text{Var} \left( X_{i,j}^{(m)} - X_{i,-j}^{(m)} \gamma \right),~ j \in [p] ,
\end{equation}
and elemental calculation yields that $\gamma_j^{(m)} = -\Omega_{-j,j}^{(m)} / \Omega_{j,j}^{(m)}$ and $\text{Var}\left( X_{i,j}^{(m)} - X_{i,-j}^{(m)} \gamma_j^{(m)} \right) = 1 / \Omega_{j,j}^{(m)}$. 
Therefore, the estimation of $\Omega^{(m)}$ can be approached through the estimation of $\gamma_j^{(m)}$'s.
Specifically, let $\widebar{X}^{(m)} = n_m^{-1} \sum_{i=1}^{n_m} X_i^{(m)}$ be the empirical estimation of the mean $\mu^{(m)}$, and denote by $\widecheck{X}^{(m)}$ the centralized data, i.e., $\widecheck{X}^{(m)} = \left( \widecheck{X}_{i,j}^{(m)} \right)_{i\in[n_m], j\in[p]} \in \bbR^{n_m \times p}$ with $\widecheck{X}_{i,j}^{(m)} := X_{i,j}^{(m)} - \widebar{X}_{j}^{(m)}$.
The coefficient vector $\gamma_j^{(m)}$ can be estimated via the node-wise Lasso procedure \citep{tibshirani1996,meinshausen2006high}:
\begin{equation}\label{eq:gamma_j_est}
\widehat{\gamma}_j^{(m) } = \arg\min_{ \gamma\in\bbR^{p-1} } \left\{ \frac{1}{2 n_m} \| \widecheck{X}_{\cdot, j}^{(m)} - \widecheck{X}_{\cdot,-j}^{(m)} \gamma \|_2^2 + \lambda_j^{(m)} \left\| \gamma \right\|_1\right\},~ j \in [p] .
\end{equation}
Then the diagonal entry $\Omega_{j,j}^{(m)}$ can be estimated by $\widehat{\Omega}_{j, j}^{(m)}$ \citep{van2014on} through
\begin{equation}\label{eq:Omega_jj_est}
\widehat{\Omega}_{j, j}^{(m)} = \frac{ n_m }{ ( \widecheck{X}_{\cdot, j}^{(m)} )\trans ( \widecheck{X}_{\cdot, j}^{(m)} - \widecheck{X}_{\cdot,-j}^{(m)} \widehat{\gamma}_j^{(m)} ) } ,
\end{equation}
and the off-diagonal vector $\Omega_{-j,j}^{(m)}$ can be estimated by
\begin{equation}\label{eq:Omega_j_est}
\widehat{\Omega}_{-j, j}^{(m)} = - \widehat{\gamma}_j^{(m)} \widehat{\Omega}_{j, j}^{(m)}.
\end{equation}

Due to the biases induced by the regularization in \eqref{eq:gamma_j_est}, debiased methods have been developed to obtain asymptotically unbiased estimators \citep[e.g.,][]{liu2013gaussian,jankova2015confidence,jankova2017honest,
ning2017general,neykov2018unified}. 
In general, the debiased estimator $\widebar{\Omega}^{(m)}$ can be represented as follows:
\begin{equation}\label{eq:Omega_debias}
\widebar{\Omega}^{(m)} = \widehat{\Omega}^{(m)} + (\widehat{\Omega}^{(m)})\trans - ( \widehat{\Omega}^{(m)} )\trans \widehat{\Sigma}^{(m)} \widehat{\Omega}^{(m)},
\end{equation}
where $ \widehat{\Sigma}^{(m)} = {n_m^{-1}} \sum_{i=1}^{n_m} (X_i^{(m)}-\widebar{X}^{(m)}) (X_i^{(m)}-\widebar{X}^{(m)})\trans = {n_m^{-1}} ( \widecheck{X}^{(m)} )\trans \widecheck{X}^{(m)}$ is the sample covariance.
It can be shown that $\widebar{\Omega}_{j, k}^{(m)}, j,k\in[p]$ is asymptotically linear with influence functions
\begin{equation}\label{eq:if_def}
\phi_{i,j,k}^{(m)} := \Omega^{(m)}_{j,k} - \Omega_{j,j}^{(m)}\Omega_{k,k}^{(m)} \epsilon_{i,j}^{(m)}\epsilon_{i,k}^{(m)} ,~i\in[n_m],
\end{equation}
where $\epsilon_{i,j}^{(m)}$ is the projection residual in \eqref{eq:gamma_j_def}, i.e.,
\begin{equation}\label{eq:epsilon_def}
\epsilon_{i,j}^{(m)} := X_{i,j}^{(m)} - \mu_j^{(m)} - (X_{i,-j}^{(m)} - (\mu_{-j}^{(m)})\trans ) \gamma_j^{(m)} .
\end{equation}
Then the asymptotic variance of $\sqrt{n_m} ( \widebar{\Omega}_{j, k}^{(m)} - \Omega_{j, k}^{(m)} )$ is given by $v_{j,k}^{(m)} := \text{Var}( \phi_{i,j, k}^{(m)} ) = \mathbb{E}( \phi_{i,j, k}^{(m)} )^2$.

To conduct statistical inference on an entry $\Omega_{j,k}^{(m)}$, typically the variance term $v_{j,k}^{(m)}$ should be estimated consistently to quantify the uncertainty of the derived estimator $\widebar{\Omega}_{j, k}^{(m)}$.
In the special case where the data distribution is jointly Gaussian, elemental calculation yields that $v_{j,k}^{(m)} = ( \Omega_{j, k}^{(m)} )^2 + \Omega_{j, j}^{(m)} \Omega_{k, k}^{(m)}$, and hence a plug-in estimator using either $\widehat{\Omega}^{(m)}$ or $\widebar{\Omega}^{(m)}$ can be directly employed to estimate $v_{j,k}^{(m)}$.
In this article, to accommodate more general distributions beyond Gaussianity, we propose to estimate $v_{j,k}^{(m)}$ through the empirical variance of the estimated influence functions:
\begin{equation}\label{eq:variance_est}
\widehat{v}_{j,k}^{(m)} = \frac{1}{n_m} \sum_{i=1}^{n_m} ( \widehat{\phi}_{i,j,k}^{(m)} )^2 ,
\end{equation}
where $\widehat{\phi}_{i,j,k}^{(m)}$ is derived via
\begin{equation}\label{eq:if_est}
\widehat{\phi}_{i,j,k}^{(m)} = \widebar{\Omega}_{j,k}^{(m)} - \widebar{\Omega}_{j,j}^{(m)} \widebar{\Omega}_{k,k}^{(m)} \widehat{\epsilon}_{i,j}^{(m)} \widehat{\epsilon}_{i,k}^{(m)}, \text{ and } \widehat{\epsilon}_{i,j}^{(m)} = \widecheck{X}_{i,j}^{(m)} - \widecheck{X}_{i,-j}^{(m)}\widehat{\gamma}_j^{(m)} .
\end{equation}

Finally, the $m$-th site transmits the sample size $n_m$ and the local estimator $\widebar{\Omega}^{(m)}$ in \eqref{eq:Omega_debias} to the central site, followed by the integration step in subsequent Section \ref{sec:method_heat}. Depending on the specific implementation of the integrative procedure, the local site may also optionally transmit the variance estimation $( \widehat{v}_{j,k}^{(m)} )_{ j,k\in[p] }$ in \eqref{eq:variance_est}.
Notably, such sharing of summary statistics in communication precludes the transfer of individual-level data, and therefore provides raw data protection in multi-source data integration.

\begin{remark}\label{remark:local_symmetry}
The constructions of debiased estimator in \eqref{eq:Omega_debias} and variance estimator in \eqref{eq:variance_est} ensure that both $\widebar{\Omega}^{(m)}$ and $( \widehat{v}_{j,k}^{(m)} )_{ j,k\in[p] }$ are symmetric, i.e., $\widebar{\Omega}_{j,k}^{(m)} = \widebar{\Omega}_{k,j}^{(m)}$ and $\widehat{v}_{j,k}^{(m)} = \widehat{v}_{k,j}^{(m)}$ for all $j,k\in[p]$.
Therefore, in practice, it is sufficient for local sites to transmit only the upper triangular entries corresponding to the index set $\{ (j,k): 1 \leq j \leq k \leq p \}$, which reduces the communication burden.
\end{remark}

\begin{remark}\label{remark:communication_efficiency}
The proposed transmission achieves communication efficiency in distributed settings, as the size of transmitted summary statistics in Algorithm \ref{alg:alg_main} is proportional to the dimension of the target parameters.
It aligns with the existing communication-efficient algorithms designed under homogeneity \citep{lee2017communication,wang2017efficient,battey2018distributed,jordan2019communication}, and shows advantage over the heterogeneous approaches \citep{cai2022individual,li2023targeting} that require the transmission of $p \times p$ second-order Hessian matrices even when the target parameter is only $p$-dimensional.
See Section \ref{sec:more_discuss_commu} of the supplement for more detailed discussions.
\end{remark}

\subsection{Heterogeneity-Adjusted Aggregating and Thresholding}\label{sec:method_heat}

To exploit the data similarity and improve the estimation accuracy, we next perform a HEterogeneity-adjusted Aggregating and Thresholding (HEAT) step based on the local estimation in the previous section to obtain integrative estimators.

We start with a reparametrization of the matrix $\Omega^{(m)}$ by decomposing it into two components: a common baseline matrix that indicates the between-sites similarity and an individual matrix that accommodates the heterogeneity.
To be specific, we write
\begin{equation}\label{eq:hete_represent}
\Omega^{(m)} = \Gamma + \Lambda^{(m)},~ m \in[M],
\end{equation}
where $\Gamma = (\Gamma_{j,k})_{j,k\in[p]}$ is the common baseline matrix, $\Lambda^{(m)} = ({\Lambda}_{j,k}^{(m)})_{j,k\in[p]}$ reflects distribution heterogeneity, and we impose additional equality constraints $\sum_{m=1}^M n_m \Lambda_{j,k}^{(m)} = 0,j,k\in[p]$ for identification.
Such representation is common in the analysis of variance. 
It follows that $\Gamma=\frac{1}{N} \sum_{m=1}^M n_m \Omega^{(m)}$ is the weighted average of $\Omega^{(m)},m\in[M]$, and that $\Lambda_{j, k}^{(1)} = \cdots = \Lambda_{j, k}^{(M)}$ implies $\Lambda_{j, k}^{(\cdot)} = 0_M$ and $\Omega_{j, k}^{(1)} = \cdots = \Omega_{j, k}^{(M)} = \Gamma_{j, k}$.
 Consequently,  the heterogeneous patterns in the space \eqref{eq:para_space_mot} are characterized by $\Lambda^{(\cdot)}$.

Based on the heterogeneity representation in \eqref{eq:hete_represent}, we propose the HEAT approach to integrate the results from local sites.
Specifically, the HEAT estimators for $\Omega^{(m)},m\in[M]$ are calculated by
\begin{equation}\label{eq:Omega_est_threshold}
\widetilde{\Omega}^{(m)} = \widetilde{\Gamma} + \widetilde{\Lambda}^{(m)}, m\in[M],
\end{equation}
where the $\widetilde{\Gamma} = (\widetilde{\Gamma}_{j,k})_{j,k\in[p]}$ and $\widetilde{\Lambda}^{(m)} = (\widetilde{\Lambda}_{j,k}^{(m)})_{j,k\in[p]}$ are given by
\begin{equation}\label{eq:Gamma_est_threshold}
\widetilde{\Gamma}_{j, k} = T_{1, \lambda_{1, j, k} }( \widebar{\Omega}_{j,k} ),
\end{equation}
\begin{equation}\label{eq:Lambda_est_threshold}
\widetilde{\Lambda}_{j, k}^{(\cdot)} = T_{2, \lambda_{2, j, k}} ( \widebar{\Omega}_{j, k}^{(\cdot)} - \widebar{\Omega}_{j,k} 1_M ),
\end{equation}
with $\widebar{\Omega}_{j, k} = \frac{1}{N}\sum_{m = 1}^M n_m \widebar{\Omega}_{j, k}^{(m)}$.
Here $T_{1, \lambda}: \bbR \rightarrow \bbR$ and $T_{2, \lambda}: \bbR^M \rightarrow \bbR^M$ are some thresholding functions, and $\lambda_{1, j, k},\lambda_{2, j, k} \geq 0$ are certain shrinkage levels.
Finally, the HEAT estimator $\widetilde{\Omega}^{(m)}$ is transmitted back to the $m$-th local site for all $m\in[M]$.

We now briefly explain the proposed HEAT method above. 
Step \eqref{eq:Omega_est_threshold} can essentially be viewed as a double shrinkage approach for the individual estimators.
By the representation \eqref{eq:hete_represent}, smaller absolute values of the components in $\widebar{\Omega}_{j, k}^{(\cdot)} - \widebar{\Omega}_{j,k} 1_M$ indicate stronger signals of homogeneity. Therefore, shrinkage towards zero is performed by $T_{2,\lambda_{2,j,k}}$ to adjust for heterogeneity.
In the meanwhile, the thresholding function $T_{1,\lambda_{1,j,k}}$ shrinks the estimator $\widebar{\Omega}_{j,k}$ to encourage the sparsity patterns of the common matrix $\Gamma$.

Next, we discuss the choices of the thresholding functions and the shrinkage levels. 
By \citet{rothman2009generalized}, a univariate thresholding function $T_{1, \lambda}$ should satisfy the following properties:
\begin{equation}\label{eq:threshold_func_univariate}
T_{1, \lambda}(x)=0 \text { if }|x| \leq \lambda, \text{ and }\left|T_{1, \lambda}(x)-x\right| \leq \lambda \text { for all } x \in \bbR.
\end{equation}
Proper functions include soft, hard, SCAD \citep{fan2001variable} and MCP \citep{zhang2010nearly} thresholding. 
For the multivariate $T_{2,\lambda}$, we generalize the univariate criteria \eqref{eq:threshold_func_univariate} and propose the following multivariate thresholding rules:
\begin{equation}\label{eq:threshold_func_multivariate}
T_{2, \lambda}(x) = 0_{M} \text{ if } \|x\| \leq \lambda, \text{ and } \left\|T_{2, \lambda}(x)-x\right\| \leq \lambda \text{ for all } x \in \bbR^M,
\end{equation}
where $\|\cdot\|$ represents some norm for the space $\bbR^M$ and will be specified later in Assumption \ref{con_thresholding_fun} of Section \ref{sec:theory}.
Proper functions for $T_{2, \lambda}$ include the multivariate versions of soft, hard, SCAD and MCP thresholding; see Section \ref{sec:addition_choice_threshold} of the supplement for detailed examples.

The shrinkage levels $\lambda_{1,j,k}$ and $\lambda_{2,j,k}$ in \eqref{eq:Gamma_est_threshold}-\eqref{eq:Lambda_est_threshold} can be determined using either universal shrinkage \citep[e.g.,][]{bickel2008covariance,rothman2009generalized,cai2012optimal} or entry-dependent adaptive shrinkage \citep[e.g.,][]{cai2011adaptive,fan2013large}.
Specifically, for universal shrinkage, the thresholds $\lambda_{1,j,k}$ and $\lambda_{2,j,k}$ remain the same across all entries $(j,k)$, while adaptive shrinkage sets distinct thresholds to account for the entry-wise volatility heteroscedasticity of $\widebar{\Omega}_{j,k}$ and $\widebar{\Omega}_{j,k}^{(\cdot)} - \widebar{\Omega}_{j,k} 1_M$ for $j,k\in[p]$.
The choice between universal and adaptive shrinkage depends on whether the local variance estimators from Section \ref{sec:method_local} are transmitted to the central site.

\begin{remark}
The HEAT approach achieves computational efficiency for joint precision matrix estimation.
As reviewed in Section \ref{sec:intro}, existing joint estimation algorithms mainly involve highly non-smooth and complex regularizations, typically leading to computationally intensive optimization procedures.
In contrast, HEAT in step \eqref{eq:Omega_est_threshold} is optimization-free for integration, which therefore renders the procedure computationally efficient at the central site.
\end{remark}

%%%%%%%%%%%%%%%%%%%%%%%%%%%%%%%%%%%%%%%%%%%%%%%%%%%%%%
%%%%%%%%%%%%%%%%%%%%%%%%%%%%%%%%%%%%%%%%%%%%%%%%%%%%%%
\section{Theoretical Properties}\label{sec:theory}
%%%%%%%%%%%%%%%%%%%%%%%%%%%%%%%%%%%%%%%%%%%%%%%%%%%%%%
%%%%%%%%%%%%%%%%%%%%%%%%%%%%%%%%%%%%%%%%%%%%%%%%%%%%%%

This section presents the statistical convergence rates for Algorithm \ref{alg:alg_main} and shows its optimality by establishing the minimax information limits for the current integrative estimation task.
To begin with, we introduce proper measures to evaluate the estimation accuracy.
For $A^{ (\cdot) } = ( A^{ (1) }, \cdots, A^{ (M) } ) \in (\bbR^{p \times p})^{\otimes M}$ and $B^{ (\cdot) } = ( B^{ (1) }, \cdots, B^{ (M) } )  \in (\bbR^{p \times p})^{\otimes M}$, define the integrative loss matrices $L_{1,r}( A^{ (\cdot) }, B^{ (\cdot) }) \in \bbR^{p \times p}$ and $L_{2,r}( A^{ (\cdot) }, B^{ (\cdot) }) \in \bbR^{p \times p}$ as
\begin{equation}\label{eq:loss_L_1r}
\left( L_{1,r}(A^{ (\cdot) },B^{ (\cdot) }) \right)_{j,k} = \| A_{j,k}^{(\cdot)} - B_{j,k}^{(\cdot)} \|_{1,w}^r , ~j,k\in[p],
\end{equation}
\begin{equation}\label{eq:loss_L_2r}
\left( L_{2,r}(A^{ (\cdot) },B^{ (\cdot) }) \right)_{j,k} = \| A_{j,k}^{(\cdot)} - B_{j,k}^{(\cdot)} \|_{2,w}^r , ~j,k\in[p],
\end{equation}
where $r \geq 1$ is a fixed constant, and the weighted $\ell_1$ norm $\| \cdot \|_{1,w}$ and weighted $\ell_2$ norm $\| \cdot \|_{2,w}$ for a vector $a\in\bbR^{M}$ are given by 
$$
\| a \|_{1,w} := \frac{1}{N} \sum_{m=1}^M n_m |a_m| \text{ and } \|a\|_{2, w} := \sqrt{ \frac{1}{N} \sum_{m=1}^M n_m a_m^2} .
$$
Therefore, \eqref{eq:loss_L_1r}-\eqref{eq:loss_L_2r} quantify the entry-wise integrative deviation between $A^{(\cdot)}$ and $B^{(\cdot)}$ under the $r$-th power of the weighted $\ell_1$ and $\ell_2$ norms, respectively. 
It is noteworthy that the sample size heterogeneity across the local sites is accommodated by the weights $\{n_m / N\}_{m\in[M]}$, and consequently there is no necessity to impose the commonly assumed balanced sample size condition to derive the convergence rates.

\subsection{Rates of Convergence for HEAT}\label{sec:theory_up}

We first introduce a few technical assumptions.

\begin{assumption}\label{con_cov_regular}
There exist constants $C_{\max},C_{\min}>0$ such that $\min_{m\in[M]} \sigma_{\min}\left( \Sigma^{(m)} \right) \geq C_{\min}$ and $\max_{m\in[M]} \max_{j\in[p]} \Sigma_{j,j}^{(m)} \leq C_{\max}$, where $\sigma_{\min}\left( \Sigma^{(m)} \right)$ denotes the minimal eigenvalue of $\Sigma^{(m)}$.
\end{assumption}

\begin{assumption}\label{con_subGaussian}
$X_{i,j}^{(m)} - \mu_j^{(m)}$ and $\epsilon_{i,j}^{(m)}$ are sub-Gaussian for all $j\in[p]$ and $m\in[M]$ with uniformly bounded sub-Gaussian norms.
\end{assumption}

\begin{assumption}\label{con_if_variance}
There exists a constant $c>0$ such that $v_{j,k}^{(m)} := \text{Var}(\phi_{i,j,k}^{(m)}) = \mathbb{E}( \phi_{i,j, k}^{(m)} )^2 \geq c$ uniformly for all $j,k\in[p]$ and $m\in[M]$, where $\phi_{i,j,k}^{(m)}$ is defined in \eqref{eq:if_def}.
\end{assumption}

\begin{assumption}\label{con_thresholding_fun}
The thresholding function $T_{1, \lambda}: \bbR \rightarrow \bbR$ satisfies the properties \eqref{eq:threshold_func_univariate}, and the function $T_{2, \lambda}: \bbR^M \rightarrow \bbR^M$ satisfies the properties \eqref{eq:threshold_func_multivariate} with norm $\| \cdot \| = \| \cdot \|_{2,w}$ on $\bbR^M$.
\end{assumption}

\begin{remark}\label{remark:condition}
Assumptions \ref{con_cov_regular}-\ref{con_if_variance} impose mild conditions on data distributions, and Assumption \ref{con_thresholding_fun} summarizes the requirements for the thresholding functions.
Specifically, Assumption \ref{con_cov_regular} is a regularity condition to prevent singular covariances and avoid diverging variances.
Assumption \ref{con_subGaussian} excludes the heavy-tailed distributions, and Assumption \ref{con_if_variance} assumes non-degenerate variances for the influence functions $\phi_{i,j,k}^{(m)}$'s.
Assumptions \ref{con_subGaussian}-\ref{con_if_variance} can be implied by Assumption \ref{con_cov_regular} if the distributions are Gaussian.
The weighted norm $\| \cdot \|_{2,w}$ for $T_{2,\lambda}$ in Assumption \ref{con_thresholding_fun} accommodates the sample size differences among the distributed sites.
Section \ref{sec:addition_choice_threshold} of the supplement presents specific examples of $T_{1,\lambda}$ and $T_{2,\lambda}$ that satisfy Assumption \ref{con_thresholding_fun}, including soft, hard, SCAD and MCP thresholding functions.
\end{remark}

Next we establish the theoretical results for the HEAT approach under the integrative losses \eqref{eq:loss_L_1r}-\eqref{eq:loss_L_2r} with various matrix norms.
We begin with several propositions to provide useful guidance on the determination of the shrinkage levels $\lambda_{1,j,k}$ and $\lambda_{2,j,k}$ in the thresholding functions \eqref{eq:Gamma_est_threshold}-\eqref{eq:Lambda_est_threshold}.
The key is to provide explicit and tight quantifications for the uncertainties of the aggregated statistics $ \widebar{\Omega}_{j,k} $ and $\| \widebar{\Omega}_{j,k}^{(\cdot)} - \widebar{\Omega}_{j,k} 1_M \|_{2,w}$. 
Recall that $s_1$ and $s_2$ are the sparsity levels in the parameter space \eqref{eq:para_space_mot}, and let $s_0 := s_1 \vee s_2$.

\begin{proposition}\label{prop:event_lambda_1jk}
Under Assumptions \ref{con_cov_regular}-\ref{con_if_variance}, assume $s_0\log(p) = o(n)$, $\log^3(p) = o(N)$, $n \geq C s_0^2 \log(p)$ for some constant $C > 0$, and $\lambda_j^{(m)} \asymp \sqrt{\log(p) / n_m}$ uniformly for $j\in[p]$ and $m\in[M]$, and define 
$$
\widebar{\lambda}_{1,j,k} = 2 \sqrt{ \| v_{j,k}^{(\cdot)} \|_{1,w} \frac{ \log(p) }{ N } } + \widebar{C}_1 \frac{s_0 M \log(p)}{N} 
$$
for some constant $\widebar{C}_1 > 0$, then the event $\bigcap_{j,k\in[p]} \left\{ | \widebar{\Omega}_{j,k} - \Gamma_{j,k} | \leq \widebar{\lambda}_{1,j,k} \right\}$ holds with probability tending to 1.
\end{proposition}

\begin{proposition}\label{prop:event_lambda_2jk}
Under Assumptions \ref{con_cov_regular}-\ref{con_if_variance}, assume $s_0\log(p) = o(n)$, $n \geq C s_0^2 \log(p)$ for some constant $C > 0$, and $\lambda_j^{(m)} \asymp \sqrt{\log(p) / n_m}$ uniformly for $j\in[p]$ and $m\in[M]$, and define
$$
\widebar{\lambda}_{2,j,k} = \sqrt{ \frac{ \| v_{j, k}^{(\cdot)} \|_1 + ( 2 \sqrt{2} + \widebar{\delta} ) \| v_{j, k}^{(\cdot)} \|_2 \sqrt{ \log (p) } + ( 4 + \widebar{\delta} ) \| v_{j, k}^{(\cdot)} \|_{\infty} \log (p) }{ N } } + \widebar{C}_2 \frac{s_0 \sqrt{M} \log(p)}{\sqrt{nN}} + \frac{\tau}{ \sqrt{N} }
$$
and $\tau = \widebar{C}_3 ( n_{\max}/n ) M^{5/2} \log(p) / n^{1/2}$ for some constants $\widebar{C}_2,\widebar{C}_3 >0$,
then for any constant $\widebar{\delta} > 0$, the event
$\bigcap_{j,k\in[p]} \left\{ \| \widebar{\Omega}_{j,k}^{(\cdot)} - \widebar{\Omega}_{j,k} 1_M - \Lambda_{j,k}^{(\cdot)} \|_{2,w} \leq \widebar{\lambda}_{2,j,k} \right\}$ holds with probability tending to 1.
\end{proposition}

Propositions \ref{prop:event_lambda_1jk}-\ref{prop:event_lambda_2jk} characterize the stochastic volatility of $\widebar{\Omega}_{j,k}$ and $\widebar{\Omega}_{j,k}^{(\cdot)} - \widebar{\Omega}_{j,k} 1_M$.
As a result, $|\widebar{\Omega}_{j,k}| \leq \widebar{\lambda}_{1,j,k}$ and $\|\widebar{\Omega}_{j,k}^{(\cdot)} - \widebar{\Omega}_{j,k} 1_M\|_{2,w} \leq \widebar{\lambda}_{2,j,k}$ respectively indicate that the underlying signals $\Gamma_{j,k}$ and $\Lambda_{j,k}^{(\cdot)}$ are weak, allowing for thresholding $\widebar{\Omega}_{j,k}$ or $\widebar{\Omega}_{j,k}^{(\cdot)} - \widebar{\Omega}_{j,k} 1_M$ to exactly zero. 
The variance terms $\|v_{j,k}^{(\cdot)}\|_{1,w}$, $\|v_{j,k}^{(\cdot)}\|_1, \|v_{j,k}^{(\cdot)}\|_2, \|v_{j,k}^{(\cdot)}\|_\infty$ in $\widebar{\lambda}_{1,j,k}$ and $\widebar{\lambda}_{2,j,k}$ capture entry-dependent volatility.
% Although these terms can be replaced by a universal constant for universal shrinkage levels, the proposed thresholds adapt to heteroscedasticity and provide practical guidance on shrinkage level selection.
The remaining terms, $s_0 M \log(p) / N$ in Proposition \ref{prop:event_lambda_1jk} and $s_0 \sqrt{M} \log(p) / \sqrt{nN}$ in Proposition \ref{prop:event_lambda_2jk}, represent higher-order errors of the local estimators \eqref{eq:Omega_debias}.

The additional term $\tau/\sqrt{N}$ in Proposition \ref{prop:event_lambda_2jk} arises from the Chi-square approximation applied to the weighted sum-of-square quantity $\| \widebar{\Omega}_{j,k}^{(\cdot)} - \widebar{\Omega}_{j,k} 1_M - \Lambda_{j,k}^{(\cdot)} \|_{2,w}$ as presented in Lemma \ref{lemma:ss_chis_bound} of the supplement.
It is negligible if $\tau^2 = o( M \vee \log(p) )$, %which is implied by $M^2 \log(p) = o \left( n n_{\max}^{-1} \sqrt{ \left( 1 \vee \frac{\log(p)}{M} \right) n } \right)$.
and in the subsequent analysis this condition is assumed for simplification and clarity.
We note that this is a technical assumption that may be relaxed with further theoretical refinement; see Section \ref{sec:more_discuss_chi} of the supplement for more detailed discussions.
A key advantage of our methodology is that, unlike typical concentration results for quadratic forms or polynomials, the Chi-square approximation developed here provides both explicit constant and tight convergence rate, which facilitates both practical implementation and theoretical error analysis.

%In practice, the variances are unknown and we can replace them by the empirical estimator $\widehat{v}_{j,k}^{(m)}$ proposed in \eqref{eq:variance_est} of Section \ref{sec:method_local}.
In practice, as stated in Section \ref{sec:method}, the determination of shrinkage levels depends on whether the local variance estimators $(\widehat{v}_{j,k}^{(m)})_{j,k,m}$ are transmitted from the local sites to the central site.
For the case where the variance estimators are available at the central site and are used to construct the entry-dependent thresholds based on Propositions \ref{prop:event_lambda_1jk}-\ref{prop:event_lambda_2jk}, the following Proposition \ref{prop:variance_est_bound} establishes the uniform rate of convergence for the variance estimation.

\begin{proposition}\label{prop:variance_est_bound}

Under Assumptions \ref{con_cov_regular}-\ref{con_subGaussian}, if $s_0\log(p) = o(n)$, $n \geq C \log(p) ( s_0 \vee \log(p) )^2 $ for some constant $C>0$, and $\lambda_j^{(m)} \asymp \sqrt{\log(p) / n_m}$ uniformly for $j\in[p]$ and $m\in[M]$, then the variance estimator $\widehat{v}_{j,k}^{(m)}$ in \eqref{eq:variance_est} satisfies
$$
| \widehat{v}_{j,k}^{(m)} - v_{j,k}^{(m)} | \lesssim_\pr \max\left( a_n^{(m)}, (a_n^{(m)})^2 \right)
$$
uniformly for $j,k\in[p]$ and $m\in[M]$, where $a_n^{(m)} = \log(p\vee n_{\max}) \sqrt{ \left( s_0 \vee \log(p) \right) / n_m}$.

\end{proposition}

%By Propositions \ref{prop:event_lambda_1jk}-\ref{prop:variance_est_bound}, for a small constant $\delta >0$, the shrinkage levels $\lambda_{1,j,k}$ and $\lambda_{2,j,k}$ in \eqref{eq:Gamma_est_threshold}-\eqref{eq:Lambda_est_threshold} can be determined by 
%\begin{equation}\label{eq:lambda_1jk}
%\lambda_{1,j,k} = (2+\delta) \sqrt{ \| \widehat{v}_{j,k}^{(\cdot)} \|_{1,w} \frac{\log(p)}{N}} + \widebar{C}_1 \frac{s_0 M \log(p)}{N} ,
%\end{equation}
%\begin{equation}\label{eq:lambda_2jk}
%\lambda_{2,j,k} = (1 + \delta ) \sqrt{ \frac{ \|\widehat{v}_{j, k}^{(\cdot)} \|_1 + 2\sqrt{2} \| \widehat{v}_{j, k}^{(\cdot)} \|_2 \sqrt{ \log (p)} + 4 \| \widehat{v}_{j, k}^{(\cdot)} \|_{\infty} \log (p) }{N} } + \widebar{C}_2 \frac{ \sqrt{M} s_0 \log(p)}{ \sqrt{nN} } .
%\end{equation}
Based on the results in Propositions \ref{prop:event_lambda_1jk}-\ref{prop:variance_est_bound}, the shrinkage levels $\lambda_{1,j,k}$ and $\lambda_{2,j,k}$ in HEAT steps \eqref{eq:Gamma_est_threshold}-\eqref{eq:Lambda_est_threshold} can be specified either universally, applying a uniform threshold across all entries:
\begin{equation}\label{eq:lambda_1jk_2jk_univ}
\begin{aligned}
\lambda_{1,j,k} = & \lambda_{1} \asymp \sqrt{ \frac{\log(p)}{N}} + \frac{s_0 M \log(p)}{N} ,\\
\lambda_{2,j,k} = & \lambda_{2} \asymp  \sqrt{ \frac{ M + \log (p) }{N} } + \frac{ \sqrt{M} s_0 \log(p)}{ \sqrt{nN} } ,
\end{aligned}
\end{equation}
or adaptively using entry-dependent thresholds that incorporate local variance estimates:
\begin{equation}\label{eq:lambda_1jk_2jk_adap}
\begin{aligned}
\lambda_{1,j,k} = & (2+\delta) \sqrt{ \| \widehat{v}_{j,k}^{(\cdot)} \|_{1,w} \frac{\log(p)}{N}} + \widebar{C}_1 \frac{s_0 M \log(p)}{N} , \\
\lambda_{2,j,k} = & (1 + \delta ) \sqrt{ \frac{ \|\widehat{v}_{j, k}^{(\cdot)} \|_1 + 2\sqrt{2} \| \widehat{v}_{j, k}^{(\cdot)} \|_2 \sqrt{ \log (p)} + 4 \| \widehat{v}_{j, k}^{(\cdot)} \|_{\infty} \log (p) }{N} } + \widebar{C}_2 \frac{ \sqrt{M} s_0 \log(p)}{ \sqrt{nN} } ,
\end{aligned}
\end{equation}
where $\delta >0$ is a small constant.
With these two specifications for the shrinkage levels, the subsequent theorems establish the rates of convergence for the HEAT estimator $\widetilde{\Omega}^{(\cdot)}$ in \eqref{eq:Omega_est_threshold} under the loss matrices $L_{1,r}$ and $L_{2,r}$ in \eqref{eq:loss_L_1r}-\eqref{eq:loss_L_2r}.
Specifically, Theorem \ref{thm:Omega_est_integrative_l} evaluates the integrative estimation losses under the general matrix $l$-norm with $l\in[1,\infty]$, while Theorem \ref{thm:Omega_est_integrative_F} investigates the matrix Frobenius norm, which essentially vectorizes the loss matrices and measures the accuracy through vector $\ell_2$ norm.

\begin{theorem}\label{thm:Omega_est_integrative_l}
Under Assumptions \ref{con_cov_regular}-\ref{con_thresholding_fun}, if $s_0\log(p) = o(n)$, $\log^3(p) = o(N)$, $n \geq C s_0^2 \log(p)$ for some constant $C > 0$, $\tau^2 = o ( M \vee \log(p) )$ and $\lambda_j^{(m)} \asymp \sqrt{\log(p) / n_m}$ uniformly for $j\in[p]$ and $m\in[M]$, and additionally assume $\log^2(p \vee n_{\max}) (s_0 \vee \log(p)) = o(n)$ for the shrinkage levels in \eqref{eq:lambda_1jk_2jk_adap}, then for any $r \geq 1$ and $l\in[1,\infty]$, we have 
$$
\| L_{1,r}(\widetilde{\Omega}^{(\cdot)}, \Omega^{(\cdot)}) \|_l \lesssim_\pr s_1 (\varphi_1)^{r} + s_2 (\varphi_2)^{r} ,
$$
$$
\| L_{2,r}(\widetilde{\Omega}^{(\cdot)}, \Omega^{(\cdot)}) \|_l \lesssim_\pr s_1 (\varphi_1)^{r} + s_2 (\varphi_2)^{r} ,
$$
where $\varphi_1 = \sqrt{\frac{ \log(p) }{ N }} + \frac{ s_0 M \log(p)}{N}$ and $\varphi_2 = \sqrt{\frac{M + \log(p)}{N}} + \frac{ s_0 \sqrt{M} \log(p)}{ \sqrt{n N} }$.
\end{theorem}

\begin{theorem}\label{thm:Omega_est_integrative_F}
Under the conditions in Theorem \ref{thm:Omega_est_integrative_l}, we have
$$
\frac{1}{p} \| L_{1,r}(\widetilde{\Omega}^{(\cdot)}, \Omega^{(\cdot)}) \|_{\text{F}}^2 \lesssim_\pr s_1 (\varphi_1)^{2r} + s_2 (\varphi_2)^{2r} ,
$$
$$
\frac{1}{p}\| L_{2,r}(\widetilde{\Omega}^{(\cdot)}, \Omega^{(\cdot)}) \|_{\text{F}}^2 \lesssim_\pr s_1 (\varphi_1)^{2r} + s_2 (\varphi_2)^{2r} .
$$
\end{theorem}

The convergence rates in Theorem \ref{thm:Omega_est_integrative_l} consist of two components: $s_1 (\varphi_1)^{r}$ and $s_2 (\varphi_2)^{r}$, which respectively reflect the estimation errors under the homogeneity and the heterogeneity. 
The same decomposition and interpretation also apply to Theorem \ref{thm:Omega_est_integrative_F}. 
Note that both $\varphi_1$ and $\varphi_2$ consist of two terms.
The terms $\sqrt{ \frac{ \log(p) }{ N } }$ and $\sqrt{\frac{M + \log(p)}{N}}$ are the leading statistical errors for integration, while $\frac{ s_0 M \log(p)}{N}$ and $\frac{ \sqrt{M} s_0 \log(p)}{ \sqrt{n N} }$ are the higher-order errors induced by individual estimation, representing the additional costs of distributed data aggregation.
Consequently, the higher-order errors will be dominated by the leading error terms, i.e.,
\begin{equation}\label{eq:rate_varphi_psi}
\varphi_1 \asymp \psi_1 := \sqrt{\frac{ \log(p) }{ N }} \text{ and  } \varphi_2 \asymp \psi_2 := \sqrt{\frac{M + \log(p)}{N}} ,
\end{equation}
if $\frac{ M^2 }{ N } \lesssim {( s_0^2 \log(p) )}^{-1}$ and $s_0^2\log^2(p) \lesssim ( 1 \vee \frac{ \log(p) }{ M } ) n$ hold, which are further guaranteed by $\frac{ M }{ n } \lesssim ( s_0^2 \log(p) )^{-1}$.
These restrictions imply that the number of distributed sites can not be too large, and similar requirements also appear in the distributed algorithms under homogeneity \citep[e.g.,][]{zhang2015divide,lee2017communication,battey2018distributed,jordan2019communication} and heterogeneity \citep[e.g.,][]{maity2022meta,cai2022individual}.
In the following section, it will be shown that $\psi_1$ and $\psi_2$ in \eqref{eq:rate_varphi_psi} achieve the minimax optimal rates.

%%%%%%%%%%%%%%%%%%%%%%%%%%%%%%%%%%%%%%%%%%%%%%%%%%%%%%
\subsection{Minimax Lower Bounds for Integrative Estimation}\label{sec:theory_low}
%%%%%%%%%%%%%%%%%%%%%%%%%%%%%%%%%%%%%%%%%%%%%%%%%%%%%%

In this section, the minimax lower bounds are established for the integrative estimation to demonstrate the optimality of HEAT approach under various loss criteria.
Whenever there is no ambiguity, we abbreviate the parameter space $\Theta(s_1,s_2)$ defined in \eqref{eq:para_space_mot} to $\Theta$.

\begin{theorem}\label{thm:Omega_est_integrative_lower_l}
Suppose that $X_i^{(m)} \sim N(\mu^{(m)}, \Sigma^{(m)})$ for $i\in[n_m],m\in[M]$ and Assumption \ref{con_cov_regular} holds.
If $3 \leq s_1 < p/2$, $3 \leq s_2 < p/2$, $s_0^2 \log^3(p) = o(N)$, $s_2^2 M = o(N)$, $N \lesssim nM$ and $p \gtrsim N^{\beta}$ for some $\beta > 1$, then for any $r \geq 1$ and $l\in[1,\infty]$,
$$
\inf_{  \widetilde{\Omega}^{(\cdot)} } \sup_{ \Omega^{(\cdot)} \in \Theta } \mathbb{E} \| L_{1,r}(\widetilde{\Omega}^{(\cdot)}, \Omega^{(\cdot)}) \|_l \gtrsim  s_1 ( \psi_1 )^{r} + s_2 ( \psi_2 )^{r}  ,
$$
$$
\inf_{  \widetilde{\Omega}^{(\cdot)} } \sup_{ \Omega^{(\cdot)} \in \Theta } \mathbb{E}  \| L_{2,r}(\widetilde{\Omega}^{(\cdot)}, \Omega^{(\cdot)}) \|_l \gtrsim  s_1 ( \psi_1 )^{r} + s_2 ( \psi_2 )^{r}  .
$$
\end{theorem}

\begin{theorem}\label{thm:Omega_est_integrative_lower_F}
Under the conditions in Theorem \ref{thm:Omega_est_integrative_lower_l}, we have
$$
\inf_{  \widetilde{\Omega}^{(\cdot)} } \sup_{ \Omega^{(\cdot)} \in \Theta } \frac{1}{p} \mathbb{E} \| L_{1,r}(\widetilde{\Omega}^{(\cdot)}, \Omega^{(\cdot)}) \|_{\text{F}}^2 \gtrsim s_1 ( \psi_1 )^{2r} + s_2 ( \psi_2 )^{2r} ,
$$
$$
\inf_{  \widetilde{\Omega}^{(\cdot)} } \sup_{ \Omega^{(\cdot)} \in \Theta } \frac{1}{p} \mathbb{E} \| L_{2,r}(\widetilde{\Omega}^{(\cdot)}, \Omega^{(\cdot)}) \|_{\text{F}}^2 \gtrsim s_1 ( \psi_1 )^{2r} + s_2 ( \psi_2 )^{2r} .
$$
\end{theorem}

Theorems \ref{thm:Omega_est_integrative_lower_l}-\ref{thm:Omega_est_integrative_lower_F} establish the minimax lower bounds % for integrative precision matrix estimation 
under the matrix $l$-norm and Frobenius norm, respectively.
The condition $p \gtrsim N^{\beta}$ is technical and serves to quantify the difficulty of distinguishing the least favorable parameters by calculating the divergences between carefully constructed distributions.
Conditions $s_1 \geq 3$ and $s_2 \geq 3$ are also technical in constructing the least favorable distributions, and the results in Theorems \ref{thm:Omega_est_integrative_lower_l}-\ref{thm:Omega_est_integrative_lower_F} remain valid when either $s_1$ or $s_2$ equals 0, as detailed in the proofs.
Despite the extensive literature on integrative precision matrix estimation as reviewed in Section \ref{sec:literature}, to the best of our knowledge, Theorems \ref{thm:Omega_est_integrative_lower_l}-\ref{thm:Omega_est_integrative_lower_F} are among the first to establish the minimax theories.
The lower bound techniques and the matrix constructions in the proofs of Theorems \ref{thm:Omega_est_integrative_lower_l}-\ref{thm:Omega_est_integrative_lower_F} are novel and are of independent interest to the integrative precision matrix analysis under heterogeneity.

By combining Theorems \ref{thm:Omega_est_integrative_lower_l}-\ref{thm:Omega_est_integrative_lower_F} and Theorems \ref{thm:Omega_est_integrative_l}-\ref{thm:Omega_est_integrative_F}, it is shown that HEAT achieves rate-optimality under various integrative losses subject to certain restrictions on the number of distributed sites.
In the following section, we will introduce a refined HEAT method which attains enhanced estimation accuracy and optimality through iterative communications between the central and local sites.

%%%%%%%%%%%%%%%%%%%%%%%%%%%%%%%%%%%%%%%%%%%%%%%%%%%%%%
%%%%%%%%%%%%%%%%%%%%%%%%%%%%%%%%%%%%%%%%%%%%%%%%%%%%%%
\section{Iterative HEAT Estimation}\label{sec:iteration}
%%%%%%%%%%%%%%%%%%%%%%%%%%%%%%%%%%%%%%%%%%%%%%%%%%%%%%
%%%%%%%%%%%%%%%%%%%%%%%%%%%%%%%%%%%%%%%%%%%%%%%%%%%%%%

As demonstrated in Section \ref{sec:theory}, achieving optimality for HEAT requires additional constraints on the number of distributed sites due to the higher-order errors of aggregated estimators.
Moreover, even with a small or moderately large number of distributed sites, reducing these errors can enhance numerical performance in practice.
To address this, we further propose the iterative HEAT (IteHEAT) method, which iteratively refines the higher-order errors of HEAT through multi-round communications between central and local sites.
Notably, unlike the existing iterative distributed algorithms developed for homogeneous settings \citep[e.g.,][]{wang2017efficient,jordan2019communication,chen2019quantile}, addressing distributional heterogeneity presents distinct challenges and necessitates both methodological and theoretical innovations. 

In the following, we outline the proposed IteHEAT in Algorithm \ref{alg:alg_main_ite} and provide detailed descriptions and theoretical properties in Sections \ref{sec:iteration_method} and \ref{sec:iteration_theory}.

\begin{breakablealgorithm}
\caption{Iterative Distributed Estimation for Heterogeneous Precision Matrices}
\label{alg:alg_main_ite}

\begin{algorithmic}

~

\textbf{First-Round Communication:}

\begin{enumerate}[nosep]

\item {\it Sample Splitting}:

For $m\in[M]$, at the $m$-th site, do:

	\begin{enumerate}[nosep]

	\item Randomly split the samples into $H$ disjoint sets $\calI_1^{(m)}, \cdots, \calI_H^{(m)}$ such that $[n_m] = \bigcup_{h\in[H]} \calI_h^{(m)}$ and $| \calI_h^{(m)}| \asymp n_m / H $.

	\item For $h\in[H]$, compute $\widehat{\Omega}_{-h}^{(m)}$ in \eqref{eq:Omega_j_jj_est_ite} based on $\{ X_i^{(m)}, i\in \calI_{-h}^{(m)} := \bigcup_{ h^\prime \neq h } \calI_{ h^\prime }^{(m)} \}$.

	\end{enumerate}

\item {\it Initialization}:

Implement Algorithm \ref{alg:alg_main}, and obtain the first-round HEAT estimator $\widetilde{\Omega}^{(m)}$ for $m\in[M]$. Define $\widetilde{\Omega}^{(m,1)} := \widetilde{\Omega}^{(m)}$, $m\in[M]$.

\end{enumerate}

\textbf{Multi-Round Communications:}

For the iterations $t = 2,\ldots,T$, do:

\begin{enumerate}[nosep]

\item {\it Individual Local Estimation:}

For $m \in [M]$, at the $m$-th site, do:

\begin{enumerate}[nosep]

\item Cross-fitted and aggregated debiasing:
$
\widebar{\Omega}^{(m,t-1/2)} = n_m^{-1} \sum_{ h\in[H] }  |\calI_h^{(m)}| \widebar{\Omega}_h^{(m,t-1/2)} 
$, 
where 
$$
\widebar{\Omega}_h^{(m,t-1/2)} = ( \widehat{\Omega}_{-h}^{(m)} )\trans + \widetilde{\Omega}^{(m,t-1)} - ( \widehat{\Omega}_{-h}^{(m)} )\trans \widehat{\Sigma}_h^{(m)} \widetilde{\Omega}^{(m,t-1)} ,
$$
and $\widehat{\Sigma}_h^{(m)}$ is the empirical covariance matrix based on $\{ X_i^{(m)}, i\in \calI_h^{(m)} \}$.

\item Symmetrization: 
$
\widebar{\Omega}^{(m,t)} = \frac{1}{2} \left( \widebar{\Omega}^{(m,t-1/2)} + (\widebar{\Omega}^{(m,t-1/2)})\trans \right) .
$

\item Transmit $\widebar{\Omega}^{(m,t)}$ to the central site.

\end{enumerate}

\item {\it IteHEAT Estimation:}

At the central site, do:
\begin{enumerate}[nosep]

% \item Shrinkage threshold: compute the $\lambda_{1,j,k}^{(t)}$ and $\lambda_{2,j,k}^{(t)}$ based on the $( n_m )_{m\in[M]}$, $( \kappa_m )_{m\in[M]}$ and $( \widehat{v}_{j,k}^{(m_{1,2})} )_{m\in[M]}$ as well as the {\red equations }.

\item Integrative estimation via HEAT:
$$
\widetilde{\Omega}^{(m,t)} = \widetilde{\Gamma}^{(t)} + \widetilde{\Lambda}^{(m,t)} = ( \widetilde{\Gamma}_{j,k}^{(t)} )_{j,k\in[p]} + ( \widetilde{\Lambda}_{j,k}^{(m,t)} )_{j,k\in[p]},
$$
where $$
\widetilde{\Gamma}_{j,k}^{(t)} = T_{1,\lambda_{1,j,k}^{(t)}} ( \widebar{\Omega}_{j,k}^{(t)} ), 	~
\widetilde{\Lambda}_{j,k}^{(\cdot,t)} = T_{2,\lambda_{2,j,k}^{(t)}} ( \widebar{\Omega}_{j,k}^{(\cdot,t)} - \widebar{\Omega}_{j,k}^{(t)} 1_M ),
$$
and $\widebar{\Omega}^{(t)} = \frac{1}{N} \sum_{m=1}^{M} n_m \widebar{\Omega}^{(m,t)}$. %, and shrinkage levels $\lambda_{1,j,k}^{(t)}$, $\lambda_{2,j,k}^{(t)}$ in \eqref{eq:lambda1t}-\eqref{eq:lambda2t}.

\item Send $\widetilde{\Omega}^{(m,t)}$ back to the $m$-th local site for $m\in[M]$.

\end{enumerate}

\end{enumerate}

\end{algorithmic}
\end{breakablealgorithm}

%%%%%%%%%%%%%%%%%%%%%%%%%%%%%%%%%%%%%%%%%%%%%%%%%%%%%%
\subsection{IteHEAT Algorithm}\label{sec:iteration_method}
%%%%%%%%%%%%%%%%%%%%%%%%%%%%%%%%%%%%%%%%%%%%%%%%%%%%%%

IteHEAT begins with a preliminary data splitting and initialization using Algorithm \ref{alg:alg_main}, and then refines the estimates through iterative local debiasing and central integration. This approach offers insights into the statistical refinement of higher-order errors and the accuracy enhancement in integrative analysis.

First, for a fixed integer $H \geq 2$, each site splits the data into $H$ disjoint subsets, and these subsets are used to construct $H$ regularized estimators for $\Omega^{(m)}$.
Specifically, for each $m\in[M]$, the $m$-th site randomly splits the samples into $H$ disjoint folds $\calI_1^{(m)}, \cdots, \calI_H^{(m)}$ such that $[n_m] = \bigcup_{h\in[H]} \calI_h^{(m)}$ and $| \calI_h^{(m)}| \asymp n_m / H $, and we denote $\calI_{-h}^{(m)} := \bigcup_{ h^\prime \neq h } \calI_{ h^\prime }^{(m)}$.
Based on this partition, we define the data $X_h^{(m)} := X_{\calI_h^{(m)}, \cdot }^{(m)} \in \bbR^{ | \calI_h^{(m)}| \times p }$ and $X_{-h}^{(m)} := X_{ \calI_{-h}^{(m)}, \cdot }^{(m)} \in \bbR^{ (n_m - |\calI_h^{(m)}| ) \times p }$ as well as the corresponding centralized data $\widecheck{X}_h^{(m)} \in \bbR^{ |\calI_h^{(m)}| \times p }$ and $\widecheck{X}_{-h}^{(m)} \in \bbR^{ (n_m-|\calI_h^{(m)}|) \times p }$, where $(\widecheck{X}_h^{(m)})_{i,j} = (X_h^{(m)})_{i,j} - (\widebar{X}_h^{(m)})_j$ and $(\widecheck{X}_{-h}^{(m)})_{i,j} = (X_{-h}^{(m)})_{i,j} - (\widebar{X}_{-h}^{(m)})_j$ and empirical means $\widebar{X}_h^{(m)} = \frac{1}{|\calI_h^{(m)}|} \sum_{i \in \calI_h^{(m)} } X_i^{(m)}$ and $\widebar{X}_{-h}^{(m)} = \frac{1}{ |\calI_{-h}^{(m)}| } \sum_{i \in \calI_{-h}^{(m)} } X_i^{(m)}$.
Then, similar to the local estimator $\widehat{\Omega}^{(m)}$ in \eqref{eq:Omega_jj_est}-\eqref{eq:Omega_j_est}, for each fold $h\in[H]$, the regularized estimation $\widehat{\Omega}_{-h}^{(m)} \in \bbR^{p \times p}$ is constructed based on the data $X_{-h}^{(m)}$:
\begin{equation}\label{eq:Omega_j_jj_est_ite}
(\widehat{\Omega}_{-h}^{(m)} )_{j, j} = \frac{ |\calI_{-h}^{(m)}| }{ ( (\widecheck{X}_{-h}^{(m)})_{\cdot, j} )\trans ( (\widecheck{X}_{-h}^{(m)})_{\cdot, j} - (\widecheck{X}_{-h}^{(m)})_{\cdot, -j} \widehat{\gamma}_{-h,j}^{(m)} ) }, \text{ and } (\widehat{\Omega}_{-h}^{(m)})_{-j, j} = - \widehat{\gamma}_{-h,j}^{(m)} (\widehat{\Omega}_{-h}^{(m)} )_{j, j} ,
\end{equation}
where
\begin{equation}\label{eq:gamma_j_est_ite}
\widehat{\gamma}_{-h,j}^{(m)} = \arg\min_{ \gamma\in\bbR^{p-1} } \left\{ \frac{1}{2 |\calI_{-h}^{(m)}| } \| (\widecheck{X}_{-h}^{(m)})_{\cdot, j} - (\widecheck{X}_{-h}^{(m)})_{\cdot, -j} \gamma \|_2^2 + \sqrt{ \frac{ n_m }{ |\calI_{-h}^{(m)}| } } \lambda_j^{(m)}  \left\| \gamma \right\|_1\right\} .
\end{equation}
Thus, in contrast to the estimator $\widehat{\Omega}^{(m)}$ in Algorithm \ref{alg:alg_main}, the $\widehat{\Omega}_{-h}^{(m)}$ is obtained by excluding the samples in subset $\calI_h^{(m)}$.

Next, Algorithm \ref{alg:alg_main} is implemented to obtain the initial first-round HEAT estimator $\widetilde{\Omega}^{(\cdot,1)} := \widetilde{\Omega}^{(\cdot)}$, followed by multi-round communications to achieve the IteHEAT estimation.
For $t \geq 2$, during the $t$-th round of communication, the $m$-th site computes the symmetrized debiased estimator by
\begin{equation}\label{eq:Omega_debias_ite}
\widebar{\Omega}^{(m,t)} = \frac{1}{2} \left( \widebar{\Omega}^{(m,t-1/2)} + (\widebar{\Omega}^{(m,t-1/2)})\trans \right),
\end{equation}
where
\begin{equation}\label{eq:Omega_debias_ite_cross}
\widebar{\Omega}^{(m,t-1/2)} =  \sum_{ h\in[H] } \frac{ |\calI_h^{(m)}| }{n_m} \widebar{\Omega}_h^{(m,t-1/2)}
\end{equation}
is the aggregated estimator across the multiple split folds and 
\begin{equation}\label{eq:Omega_debias_ite_split}
\widebar{\Omega}_h^{(m,t-1/2)} = ( \widehat{\Omega}_{-h}^{(m)} )\trans + \widetilde{\Omega}^{(m,t-1)} - ( \widehat{\Omega}_{-h}^{(m)} )\trans \widehat{\Sigma}_h^{(m)} \widetilde{\Omega}^{(m,t-1)} , ~ h\in[H].
\end{equation}
Here, $\widehat{\Sigma}_h^{(m)}$ denotes the empirical covariance matrix based on the subset $\calI_h^{(m)}$, i.e., $\widehat{\Sigma}_h^{(m)} = \frac{1}{ |\calI_h^{(m)}| } ( \widecheck{X}_h^{(m)} )\trans \widecheck{X}_h^{(m)}$.
The debiasing steps in \eqref{eq:Omega_debias_ite_cross} and \eqref{eq:Omega_debias_ite_split} mainly incorporate the idea of cross-fitting \citep{chernozhukov2018double,chernozhukov2022locally} into the  construction in \eqref{eq:Omega_debias}, and the step in \eqref{eq:Omega_debias_ite} ensures the matrix symmetrization.

Then, $\widebar{\Omega}^{(m,t)}$ is transmitted to the central site for all $m\in[M]$, followed by the HEAT estimation:
\begin{equation}\label{eq:Omega_est_threshold_ite}
\widetilde{\Omega}^{(m,t)} = \widetilde{\Gamma}^{(t)} + \widetilde{\Lambda}^{(m,t)} = ( \widetilde{\Gamma}_{j,k}^{(t)} )_{j,k\in[p]} + ( \widetilde{\Lambda}_{j,k}^{(m,t)} )_{j,k\in[p]},
\end{equation}
where
$$
\widetilde{\Gamma}_{j,k}^{(t)} = T_{1,\lambda_{1,j,k}^{(t)}} ( \widebar{\Omega}_{j,k}^{(t)} ), \text{ and } \widetilde{\Lambda}_{j,k}^{(\cdot,t)} = T_{2,\lambda_{2,j,k}^{(t)}} ( \widebar{\Omega}_{j,k}^{(\cdot,t)} - \widebar{\Omega}_{j,k}^{(t)} 1_M ),
$$
and $\widebar{\Omega}^{(t)} = \frac{1}{N} \sum_{m=1}^{M} n_m \widebar{\Omega}^{(m,t)}$.
Similar to the determination of $\lambda_{1,j,k}$ and $\lambda_{2,j,k}$ as detailed in Section \ref{sec:method_heat} and Section \ref{sec:theory_up}, either universal shrinkage or entry-dependent adaptive shrinkage levels can be applied for $\lambda_{1,j,k}^{(t)}$ and $\lambda_{2,j,k}^{(t)}$ at the $t$-th iteration.

During each iteration $t\geq 2$, $\widetilde{\Omega}^{(m,t)}$ is sent back to the $m$-th local site for $m\in[M]$.
After $T$ rounds of communication, we obtain the final integrative estimation $\widetilde{\Omega}^{(\cdot,T)}$.

\begin{remark}
In spite of the independence between $\widehat{\Sigma}_h^{(m)}$ and $\widehat{\Omega}_{-h}^{(m)}$,
the debiasing step in \eqref{eq:Omega_debias_ite_split} is not a standard cross-fitting procedure as in \citet{chernozhukov2018double,chernozhukov2022locally}, and the distinction arises because the IteHEAT estimator $\widetilde{\Omega}^{(m,t-1)}$ from the previous $(t-1)$-th round is dependent with the full local data $X^{(m)}$. 
Nevertheless, this procedure yields valid debiased estimator in \eqref{eq:Omega_debias_ite}-\eqref{eq:Omega_debias_ite_cross}, and the inclusion of $\widetilde{\Omega}^{(m,t-1)}$ serves to iteratively refine the higher-order errors of the local estimators.
\end{remark}

%%%%%%%%%%%%%%%%%%%%%%%%%%%%%%%%%%%%%%%%%%%%%%%%%%%%%%
\subsection{Theoretical Properties}\label{sec:iteration_theory}
%%%%%%%%%%%%%%%%%%%%%%%%%%%%%%%%%%%%%%%%%%%%%%%%%%%%%%

This section establishes the theoretical properties of the IteHEAT approach in Algorithm \ref{alg:alg_main_ite}.
We begin with the following additional technical assumption.

\begin{assumption}\label{con_cov_subG_ite}
$\max_{m\in[M]} \sigma_{\max} \left( \Sigma^{(m)} \right) \leq C_{\max} $, where $\sigma_{\max} \left( \Sigma^{(m)} \right)$ is the maximal eigenvalue of $\Sigma^{(m)}$, and $X_i^{(m)} - \mu^{(m)}$ is sub-Gaussian with bounded sub-Gaussian norm uniformly for $m\in[M]$.
\end{assumption}

Assumption \ref{con_cov_subG_ite} is common and mild in high-dimensional statistics literature, though it is slightly stricter than Assumptions \ref{con_cov_regular}-\ref{con_subGaussian}.
The requirement for bounded maximal eigenvalues is assumed for simplicity and clarity, and it can be relaxed to bounded sparsely restricted eigenvalues.

Next, we establish the error analysis for the IteHEAT estimators $\widetilde{\Omega}^{(\cdot,t)},t \geq 2$ in \eqref{eq:Omega_est_threshold_ite} under loss matrices \eqref{eq:loss_L_1r}-\eqref{eq:loss_L_2r}.
For the shrinkage levels $\lambda_{1,j,k}^{(t)}$ and $\lambda_{2,j,k}^{(t)}$ specified in Equations \eqref{eq:lambda_1jk_2jk_univ_ite}-\eqref{eq:lambda_1jk_2jk_adap_ite} in Section \ref{sec:theorem_56} of the supplement, the following theorems present iterative contraction rates of convergence for IteHEAT.

\begin{theorem}\label{thm:Omega_est_integrative_l_ite}
Under Assumptions \ref{con_cov_regular}-\ref{con_cov_subG_ite}, if $s_0\log(p) = o(n)$, $\log^3(p) = o(N)$, $n \geq C s_0^2 \log(p)$ for some constant $C > 0$, $\tau^2 = o ( M \vee \log(p) )$ and $\lambda_j^{(m)} \asymp \sqrt{\log(p) / n_m}$ uniformly for $j\in[p]$ and $m\in[M]$, and additionally assume $\log^2(p \vee n_{\max}) (s_0 \vee \log(p)) = o(n)$ for the shrinkage levels in Equation \eqref{eq:lambda_1jk_2jk_adap_ite}, then for $t \geq 2$, $r \geq 1$ and $l\in[1,\infty]$, we have 
$$
\| L_{1,r}(\widetilde{\Omega}^{(\cdot,t)}, \Omega^{(\cdot)}) \|_l \lesssim_\pr s_1 ( \varphi_1^{(t)} )^{r} + s_2 ( \varphi_2^{(t)} )^{r} ,
$$
$$
\| L_{2,r}(\widetilde{\Omega}^{(\cdot,t)}, \Omega^{(\cdot)}) \|_l \lesssim_\pr s_1 ( \varphi_1^{(t)} )^{r} + s_2 ( \varphi_2^{(t)} )^{r} ,
$$
where 
$$
\varphi_1^{(t)} = \varphi_1^{\star} + s_0 \sqrt{ \frac{\log(p)}{n} } \varphi_2^{\star} + \left( s_0 \sqrt{ \frac{\log(p)}{n} } \right)^{t-1} ( \varphi_1 + \varphi_2 ),
$$  
$$
\varphi_2^{(t)} = \varphi_2^{\star} + \left( s_0 \sqrt{ \frac{\log(p)}{n} } \right)^{t-1} ( \varphi_1 + \varphi_2 ) ,
$$
and $\varphi_1^{\star} = \sqrt{ \frac{ \log(p) }{ N } } + \frac{ M }{ N }$, $\varphi_2^{\star} = \sqrt{ \frac{ M + \log(p) }{ N } } + \sqrt{ \frac{ \log(p) }{ N } \frac{ s_0 \log^2(p) }{ n } }$.
\end{theorem}

\begin{theorem}\label{thm:Omega_est_integrative_F_ite}
Under the conditions in Theorem \ref{thm:Omega_est_integrative_l_ite}, we have
$$
\frac{1}{p} \| L_{1,r}(\widetilde{\Omega}^{(\cdot,t)}, \Omega^{(\cdot)}) \|_{\text{F}}^2 \lesssim_\pr s_1 ( \varphi_1^{(t)} )^{2r} + s_2 ( \varphi_2^{(t)} )^{2r} ,
$$
$$
\frac{1}{p}\| L_{2,r}(\widetilde{\Omega}^{(\cdot,t)}, \Omega^{(\cdot)}) \|_{\text{F}}^2 \lesssim_\pr s_1 ( \varphi_1^{(t)} )^{2r} + s_2 ( \varphi_2^{(t)} )^{2r} .
$$
\end{theorem}

The results for IteHEAT are analogous to those for HEAT in Theorems \ref{thm:Omega_est_integrative_l}-\ref{thm:Omega_est_integrative_F}, with $\varphi_1$ and $\varphi_2$ replaced by $\varphi_1^{(t)}$ and $\varphi_2^{(t)}$, respectively.
%Both $\varphi_1^{(t)}$ and  $\varphi_2^{(t)}$ consist of iteration-independent and iteration-dependent terms, which respectively represent the {\blue fixed statistical errors} and iteratively refined higher-order errors.
For the iteration-dependent errors in $\varphi_1^{(t)}$ and $\varphi_2^{(t)}$, the factor $s_0 \sqrt{ \log(p) / n}$ serves as the contraction rate for iterative higher-order error reduction and $\left( s_0 \sqrt{ \log(p) / n} \right)^{t-1}$ leads to the geometric convergence.
Such contraction implies that condition $n \gtrsim s_0^2 \log(p)$ is required for convergence, i.e., the local sample sizes cannot be excessively small.
Similar requirements are also present in other iterative distributed algorithms designed under homogeneity \citep{wang2017efficient,chen2019quantile,jordan2019communication,tan2022communication}.
As a result, the number of iterations $T$ with the logarithm complexity
$$
T \gtrsim 1 + \log\left( \frac{ \sqrt{ M + \log(p) } + s_0 \sqrt{M/n} \log(p) }{ \sqrt{ \log(p) } + M / \sqrt{N} } \right) / \log \left( \frac{1}{s_0} \sqrt{ \frac{n}{\log(p)} } \right) 
$$
suffices to yield the following rates for $\widetilde{\Omega}^{(\cdot,T)}$:
\begin{equation}\label{eq:rate_varphi_T}
\varphi_1^{(T)} \asymp \varphi_1^{\star} + s_0 \sqrt{ \frac{\log(p)}{n} } \varphi_2^{\star}, \text{ and } \varphi_2^{(T)} \asymp \varphi_2^{\star} .
\end{equation} 
In other words, the iterative refined higher-order errors induced by aggregating local estimators are dominated by the fixed statistical errors after logarithm rounds of communication between the central and local sites.

Notably, the convergence rates in \eqref{eq:rate_varphi_T} match the optimal rates established in Theorems \ref{thm:Omega_est_integrative_lower_l}-\ref{thm:Omega_est_integrative_lower_F}, up to additional terms $\frac{M}{N}$, $\sqrt{ \frac{ \log(p) }{ N } \frac{ s_0 \log^2(p) }{ n } }$ and $s_0 \sqrt{ \frac{\log(p)}{n} } \varphi_2^{\star}$.
The two terms $\frac{M}{N}$ and $\sqrt{ \frac{ \log(p) }{ N } \frac{ s_0 \log^2(p) }{ n } }$ arise from estimating the individual means $\widebar{X}^{(m)}, m\in[M]$ for data centralization, which relies only on the local samples.
They get dominated by $\psi_1$ and $\psi_2$ under the mild conditions $M \lesssim \sqrt{ N \log(p) }$ (or $M \lesssim n\log(p)$) and $s_0 \log^2(p) \lesssim n$.
Moreover, these data centralization errors would vanish if the means are known (e.g., $\mu^{(m)} = 0$ for $m\in[M]$), or become negligible if the means are assumed to be homogeneous across distributed sites and are estimated using the pooled empirical means.

The remaining error $s_0 \sqrt{ \frac{\log(p)}{n} } \varphi_2^{\star}$ represents the cost for estimating homogeneous components via iterative aggregation in the presence of heterogeneity. 
It is dominated by the leading error $\psi_1$ if $\frac{\sqrt{M}}{n} \lesssim (s_0^2 \log(p))^{-1}$ and $M \lesssim \frac{n}{s_0^2} \vee \log(p)$, which are weaker than those required for HEAT as shown in Section \ref{sec:theory_up}.
It is also worth noting that, this error term would vanish in homogeneous setting where $s_2=0$, which highlights the distinction between distributed estimation under homogeneity versus heterogeneity.

Consequently, after a few rounds of communication, the results in Theorems \ref{thm:Omega_est_integrative_l_ite}-\ref{thm:Omega_est_integrative_F_ite} show that IteHEAT enhances the estimation accuracy and optimality compared to HEAT by leveraging iterative refinements.

%%%%%%%%%%%%%%%%%%%%%%%%%%%%%%%%%%%%%%%%%%%%%%%%%%%%%%
%%%%%%%%%%%%%%%%%%%%%%%%%%%%%%%%%%%%%%%%%%%%%%%%%%%%%%
\section{Simulation Study}\label{sec:simulation}
%%%%%%%%%%%%%%%%%%%%%%%%%%%%%%%%%%%%%%%%%%%%%%%%%%%%%%
%%%%%%%%%%%%%%%%%%%%%%%%%%%%%%%%%%%%%%%%%%%%%%%%%%%%%%

In this section, we evaluate the simulation performance of the proposed HEAT and IteHEAT approaches, and compare them with several joint estimation procedures that have access to the full distributed datasets as oracle baselines.
The data generation procedure is detailed in Section \ref{sec:addition_data_generation} of the supplement and is briefly described as follows. 
To construct $\Omega^{(\cdot)}$, two graph structures are considered: the Erd\"os-R\'enyi random graph and the banded graph, and the heterogeneity pattern is controlled by a parameter $\texttt{hete\_ratio} \in [0,1]$, which approximately represents the ratio $\frac{s_2}{s_1+s_2}$.
Then we generate $X_i^{(m)} \sim \text{N}(0, (\Omega^{(m)})^{-1} )$ for $i\in[n_m]$ and $m\in[M]$, where the sample sizes are independently determined by $n_m = n_0 + \lceil \frac{n_0}{25} \text{N}(0,1) \rceil$, with $n_0$ being a preassigned value.
To evaluate the estimation errors, we consider the loss $L_{1,r}$ in \eqref{eq:loss_L_1r} with matrix 1-norm $\| L_{1,r} \|_1$ and Frobenius norm $\| L_{1,r} \|_{\text{F}}^2 / p$, denoted by \texttt{L\_1} and \texttt{Frobenius} respectively. 
We present the results with $r = 1$.
All simulation results are based on 256 independent replications unless stated otherwise.

For our numerical experiments, we implement HEAT and IteHEAT using both soft and SCAD thresholding functions for illustration.
Note that HEAT estimation is essentially the first-round output of IteHEAT. 
Therefore, we present the performance of the IteHEAT estimators $\widetilde{\Omega}^{(\cdot,t)}$ for $t = 0, 1, 2, \ldots,10$, where $\widetilde{\Omega}^{(\cdot,1)}$ represents the HEAT estimator, and $\widetilde{\Omega}^{(\cdot,0)} := \widehat{\Omega}^{(\cdot)}$ represents the individual local estimator in \eqref{eq:Omega_jj_est}-\eqref{eq:Omega_j_est}. 
We refer to the corresponding methods as \texttt{HEAT \& IteHEAT\_Soft} and \texttt{HEAT \& IteHEAT\_SCAD}.
In addition, we evaluate IteHEAT that employs a data-driven stopping rule for iterations, which is described in Section \ref{sec:addition_num_iteration} of the supplement, and we denote the methods by \texttt{IteHEAT\_Soft (stop)} and \texttt{IteHEAT\_SCAD (stop)}.
The practical determination of regularization parameters $\lambda_j^{(m)},j\in[p],m\in[M]$ in Lasso procedures \eqref{eq:gamma_j_est} and \eqref{eq:gamma_j_est_ite}, as well as the shrinkage levels $\lambda_{1,j,k},\lambda_{2,j,k}$ and $\lambda_{1,j,k}^{(t)},\lambda_{2,j,k}^{(t)},t=2,3,\ldots$ in HEAT \eqref{eq:Omega_est_threshold} and IteHEAT \eqref{eq:Omega_est_threshold_ite}, is discussed in Sections \ref{sec:addition_reg_para}-\ref{sec:addition_shrink_para} of the supplement.
For both HEAT and IteHEAT, we employ the entry-dependent adaptive shrinkages, and numerical comparisons between universal and adaptive shrinkages are provided in Section \ref{sec:addition_compare_shrinkage} of the supplement.
For IteHEAT, we set the number of splits to $H = 10$; see Section \ref{sec:addition_split} of the supplement for further discussion.
In the following, we set $n_0 = 400, p\in\{100,200\}$ and $ M\in\{5,20\}$.

\begin{figure}
	\centering
    \includegraphics[scale=0.45]{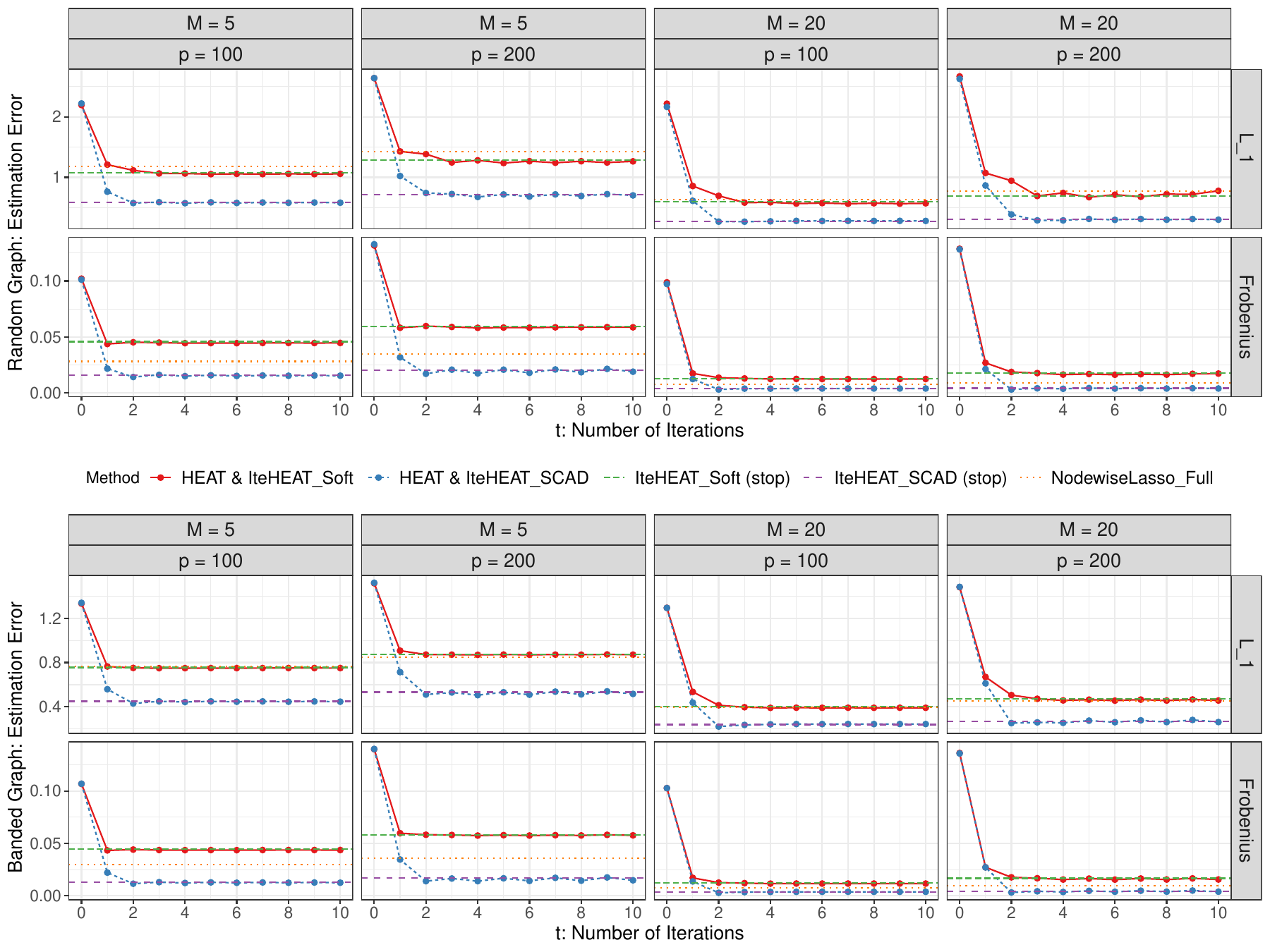}
    \caption{\small Comparisons of \texttt{HEAT \& IteHEAT}, \texttt{IteHEAT (stop)} and \texttt{NodewiseLasso\_Full} under homogeneous settings where $\texttt{hete\_ratio} = 0$.}
    \label{fig:homo}
\end{figure}
We first examine the homogeneous settings where $\texttt{hete\_ratio}=0$, indicating that $s_2 = 0$ and all datasets share the same distribution.
We compare \texttt{HEAT \& IteHEAT} and \texttt{IteHEAT (stop)} with a competing oracle method \texttt{NodewiseLasso\_Full}, which estimates the homogeneous precision matrix using the node-wise Lasso method introduced in Section \ref{sec:method_local} based on the full dataset $X_{i}^{(m)}, i\in[n_m],m\in[M]$.
The results are displayed in Figure \ref{fig:homo}.
Overall, \texttt{HEAT \& IteHEAT\_Soft} is comparable to \texttt{NodewiseLasso\_Full}, while \texttt{HEAT \& IteHEAT\_SCAD} demonstrates significant advantages over \texttt{HEAT \& IteHEAT\_Soft} and \texttt{NodewiseLasso\_Full}.
%Specifically, when $M = 5$, \texttt{HEAT \& IteHEAT\_SCAD} outperforms \texttt{NodewiseLasso\_Full} even when $t = 1$, while when $M = 20$, a small number of iterations (e.g., $t = 2 \text{ or }3$) is sufficient for \texttt{HEAT \& IteHEAT\_SCAD} to achieve high estimation accuracy.
Furthermore, the performance of \texttt{IteHEAT (stop)} closely matches the best iteration of \texttt{HEAT \& IteHEAT}, which validates the effectiveness of our proposed stopping criterion.
These results indicate that the proposed HEAT and IteHEAT exhibit competitive performance in homogeneous settings, demonstrating the validity and statistical accuracy in distributed integration.

\begin{figure}
	\centering
    \includegraphics[scale=0.45]{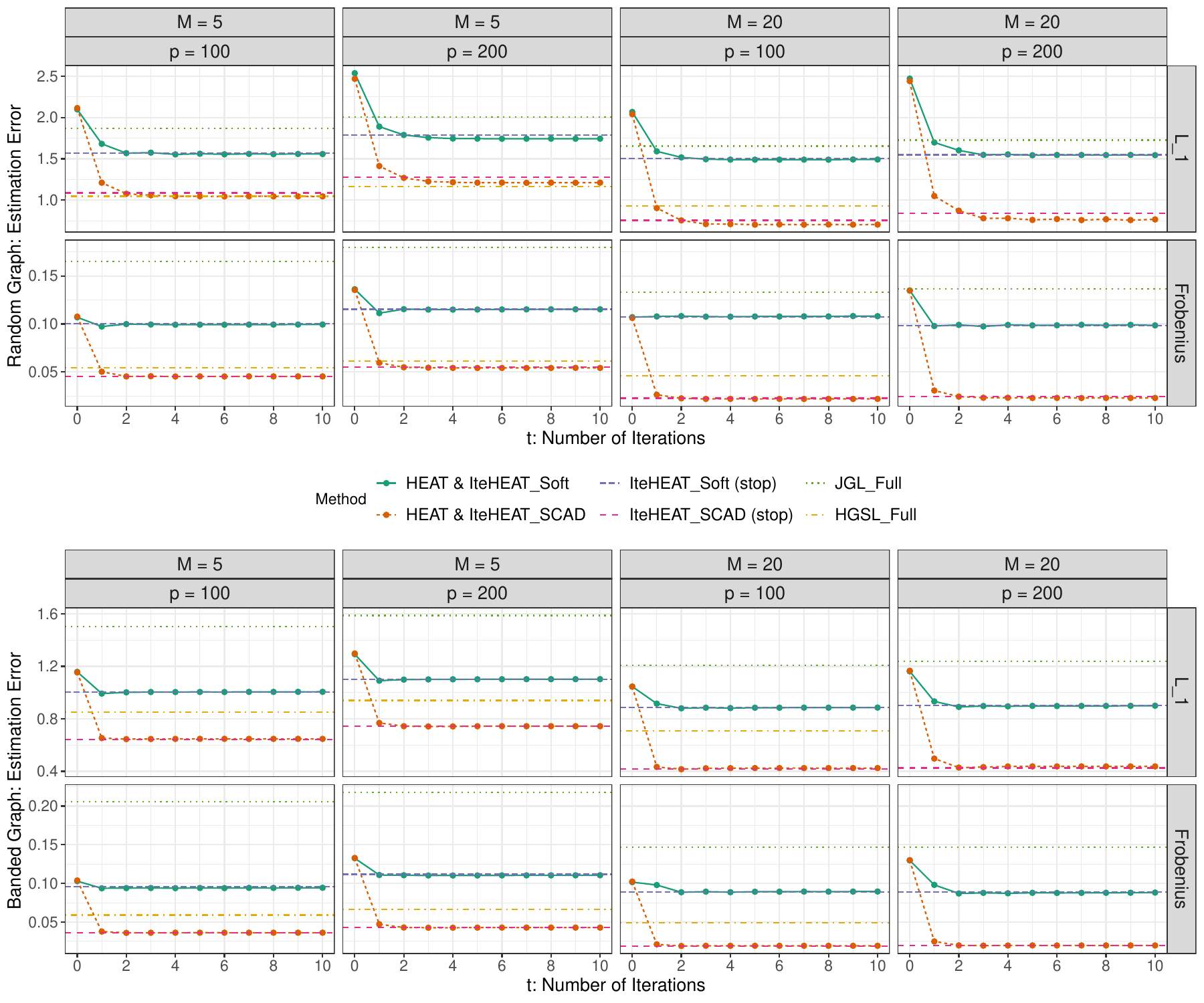}
    \caption{\small Comparisons of \texttt{HEAT \& IteHEAT}, \texttt{IteHEAT (stop)}, \texttt{JGL\_Full} and \texttt{HGSL\_Full} under completely heterogeneous settings where $\texttt{hete\_ratio} = 1$.}
    \label{fig:hete_1}
\end{figure}
Next, we compare the proposed methods with two alternative approaches under heterogeneity: Joint Graphical Lasso in \citet{danaher2014joint} and Heterogeneous Group Square-Root Lasso in \citet{ren2019tuning}, denoted by \texttt{JGL\_Full} and \texttt{HGSL\_Full}, respectively. 
Both methods utilize the full distributed datasets and employ the group Lasso penalty for estimation under shared supports.
Thus, to ensure a fair comparison, we focus on the completely heterogeneous settings where $\texttt{hete\_ratio} = 1$. 
For \texttt{JGL\_Full}, the regularization involves two parameters: $\lambda_1$ and $\lambda_2$. 
The selection of $\lambda_2$ follows the same criterion \eqref{eq:joint_criterion} as HEAT and IteHEAT described in Section \ref{sec:addition_shrink_para} of the supplement, while $\lambda_1$ is set to 0 to suit the current settings; see \citet{danaher2014joint} for more details.
For \texttt{HGSL\_Full}, the regularization parameter is determined by the tuning-free simulation strategy suggested in \citet{ren2019tuning}.
However, due to high computational costs, the performance of \texttt{HGSL\_Full} is evaluated based on only 64 replications for $(M,p)=(20,100)$, and it is not implemented for $(M,p)=(20,200)$.
The results are displayed in Figure \ref{fig:hete_1}.
It is clear that \texttt{JGL\_Full} exhibits inferior performance compared to all the other methods.
While there is a noticeable gap between \texttt{HEAT \& IteHEAT\_Soft} and \texttt{HGSL\_Full}, the proposed \texttt{HEAT \& IteHEAT\_SCAD} significantly outperforms \texttt{HGSL\_Full}, even with $t = 1$ in most settings.
Besides, consistent with the results in homogeneous settings, \texttt{IteHEAT (stop)} achieves performance comparable to the fully iterated \texttt{HEAT \& IteHEAT}.
These results further highlight the advantage of HEAT and IteHEAT in terms of statistical accuracy and flexibility in modelling heterogeneity.
%It is worth noting that the observed gap between our methods and HGSLtune\_Full can be attributed to the introduction of a common matrix $\Gamma$ in the representation \eqref{eq:hete_represent}.
%This matrix is not tailored specifically for the current fully heterogeneous setting, resulting in additional estimation errors.
%Nevertheless, HEAT and IteHEAT offer more flexibility in modelling heterogeneity beyond the considered scenarios.

\section{Real Data Analysis}\label{sec:real_data}

In this section, we explore the version 8 bulk tissue RNA-seq count data in Genotype-Tissue Expression (GTEx) project. 
This dataset collects samples from 54 tissues across nearly 1000 donors, and can be accessed publicly at \url{https://www.gtexportal.org}. 
We focus on nine tissues with the most frequently collected samples \citep{gtex2020gtex}: Adipose-Subcutaneous, Lung, Muscle-Skeletal, Skin-Sun Exposed, Skin-Not Sun Exposed, Thyroid, Nerve-Tibial, Artery-Tibial and Whole Blood, and then employ our proposed methodology to conduct differential analysis for expression networks.
Gender differences in immune diseases are widely acknowledged  \citep{moroni2012geoepidemiology,ngo2014gender}. 
Thus, we investigate gender heterogeneity within each tissue separately, focusing on the gene set related to antigen processing and presentation (GO:0019882), which contains 117 genes.
Consequently, we have $M = 2$ across tissues. 
The sample sizes in male datasets range from 395 to 543, while those in female datasets range from 183 to 260.

\begin{figure}
	\centering
    \includegraphics[scale=0.5]{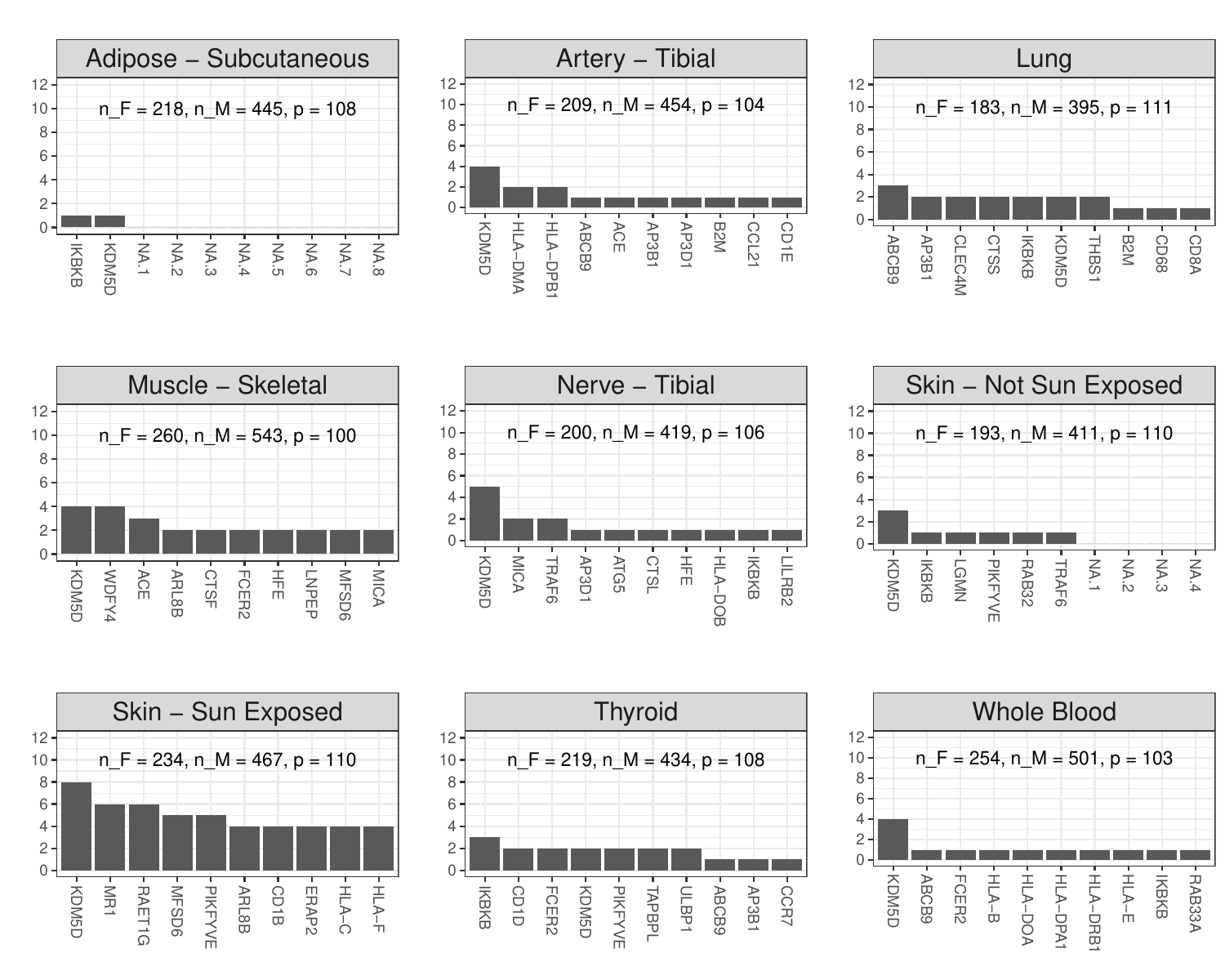}
    \caption{\small Top ten genes showing highest heterogeneous connectivity in each tissue. The y-axis shows the number of heterogeneous edges. The variables n\_F, n\_M, and p represent the numbers of female samples, male samples and remaining genes in each tissue, respectively.}
    \label{fig:graph_degree}
\end{figure}

For each tissue analysis, we employ edgeR \citep{robinson2010edger} to preprocess the RNA-seq count datasets. This includes low-expressed gene filtering, normalization, and log-counts-per-million transformation.
After preprocessing, the numbers of remaining genes range from 100 to 111.
Due to the limited number of integrated datasets ($M = 2$), we implement HEAT with SCAD thresholding and entry-dependent shrinkage levels separately for each tissue. 
We then extract estimated network edges that exhibit gender heterogeneity.
The top ten genes showing the highest connectivity within heterogeneous edges are illustrated in Figure \ref{fig:graph_degree}.
The results reveal that the gene KDM5D exhibits the highest degree of heterogeneity in connectivity. 
Additionally, heterogeneous associations related to the HLA gene family are observed in several tissues.
Notably, KDM5D, located on the Y chromosome, is recognized as a tumor suppressor \citep{ohguchi2022diverse}, and HLA genes are associated with various immune diseases \citep{ngo2014gender}. 
These findings of gender-related genetic association heterogeneity support the validity of the proposed methodology.

\section{Conclusions and Discussions}\label{sec:discuss}

In this article, we focus on distributed integrative estimation for high-dimensional precision matrices under heterogeneity and introduce the HEAT and IteHEAT methods.
These proposals achieve both communication and computation efficiency while maintaining the statistical optimality. 
More importantly, the ideas underlying HEAT and IteHEAT are not limited to precision matrix estimation alone, and can be extended to various models with unbiased local estimators that can be facilitated by modern inference methods and theories \citep[e.g.,][]{van2014on,chernozhukov2018double,chernozhukov2022locally}.

In conclusion, we raise several problems for future research. 
First, the IteHEAT in Algorithm \ref{alg:alg_main_ite}, which enhances the accuracy and optimality of HEAT in Algorithm \ref{alg:alg_main} through iterative higher-order refinement, requires a restriction on the number of local sample sizes to guarantee the convergence contraction in multi-round communications. 
Although this requirement is also present in other distributed methods under homogeneity \citep{wang2017efficient,jordan2019communication}, developing algorithms free of such restriction is a valuable problem for distributed data integration.

Second, while the proposed HEAT and IteHEAT enjoy communication efficiency as discussed in Remark \ref{remark:communication_efficiency} and Section \ref{sec:more_discuss_commu} of the supplement, adapting them for settings with exact communication constraints remains an open problem.
To our knowledge, existing works on communication-constrained sparse estimation have been limited to homogeneous Gaussian mean models \citep{garg2014communication,braverman2016communication,han2021geometric,acharya2023unified}.
Extension from homogeneous sparse mean models to heterogeneous sparse matrix models can be non-trivial and is left as future research.

Third, HEAT and IteHEAT ensure the privacy protection of individual-level data by sharing only summary statistics, which aligns with \citet{kohane2021what,duan2022heterogeneity,zhou2022global}.
Given the recent advances in areas like differential privacy, incorporating quantitative privacy-preserving mechanisms into the developed distributed learning algorithms would also be of interest.

Last, the current heterogeneity modelling in HEAT and IteHEAT may be sensitive to highly adversarial or contaminated datasets. 
Therefore, developing robust methodologies for distributed integration warrants further exploration.

\bibliographystyle{apalike}
\bibliography{ref}

\end{document}